\documentclass[10pt,aps,twocolumn,prd,superscriptaddress,showpacs,nofootinbib,floatfix,preprintnumbers]{revtex4-1}
\usepackage{graphics,graphicx}
\usepackage{amsmath, amssymb}
\usepackage{multirow}
\usepackage{color}
\usepackage{verbatim}
\usepackage{hyperref}
\usepackage[normalem]{ulem}  
\usepackage{ulem}
\usepackage{color}
\usepackage{notoccite}
\usepackage{cancel}
\usepackage{mathtools}
\usepackage{epstopdf}

\renewcommand\sout{\bgroup \color{blue} \ULdepth=-.5ex \ULset}
\def\slashchar#1{\setbox0=\hbox{$#1$}  
\dimen0=\wd0     
\setbox1=\hbox{/} \dimen1=\wd1  
\ifdim\dimen0>\dimen1   
\rlap{\hbox to \dimen0{\hfil/\hfil}} 
#1     
\else     
\rlap{\hbox to \dimen1{\hfil$#1$\hfil}} 
/      
\fi}

\newcommand{\dd}{\mathrm{d}}

\begin{document}

   \title{Towards a unified equation of state for multi-messenger astronomy}
   \date{\today}
   \author{Micha\l{} Marczenko}
   \email{michal.marczenko@uwr.edu.pl}
   \affiliation{Institute of Theoretical Physics, University of Wroclaw, PL-50204 Wroclaw, Poland}
   \author{David Blaschke}
   \affiliation{Institute of Theoretical Physics, University of Wroclaw, PL-50204 Wroclaw, Poland}
   \affiliation{Bogoliubov Laboratory of Theoretical Physics, Joint Institute for Nuclear Research, 141980 Dubna, Russia}
   \affiliation{National Research Nuclear University, 115409 Moscow, Russia}
   \author{Krzysztof Redlich}
   \affiliation{Institute of Theoretical Physics, University of Wroclaw, PL-50204 Wroclaw, Poland}
   \affiliation{Theoretical Physics Department, CERN, CH-1211 Gen\`eve 23, Switzerland}
   \author{Chihiro Sasaki}
   \affiliation{Institute of Theoretical Physics, University of Wroclaw, PL-50204 Wroclaw, Poland}

   \preprint{CERN-TH-2020-061}

\begin{abstract}
  We aim to present a first step in developing a benchmark equation-of-state (EoS) model for multi-messenger astronomy that unifies the thermodynamics of quark and hadronic degrees of freedom. A Lagrangian approach to the thermodynamic potential of quark-meson-nucleon (QMN) matter was used. In this approach, dynamical chiral-symmetry breaking is described by the scalar mean-field dynamics coupled to quarks and nucleons and their chiral partners, whereby its restoration occurs in the hadronic phase by parity doubling, as well as in the quark phase. Quark confinement was achieved by an auxiliary scalar field that parametrizes a dynamical infrared cut-off in the quark sector, serving as an ultraviolet cut-off for the nucleonic phase space. The gap equations were solved for the isospin-symmetric case, as well as for neutron star (NS) conditions. We also calculated the mass-radius (MR) relation of NSs and their tidal deformability (TD) parameter. The obtained EoS is in accordance with nuclear matter properties at saturation density and with the flow constraint from heavy ion collision experiments. For isospin-asymmetric matter, a sequential occurrence of light quark flavors is obtained, allowing for a mixed phase of chirally-symmetric nucleonic matter with deconfined down quarks. The MR relations and TDs for compact stars fulfill the constraints from the latest astrophysical observations for PSR J0740+6620, PSR J0030+0451, and the NS merger GW170817, whereby the tension between the maximum mass and compactness constraints rather uniquely fixes the model parameters. The model predicts the existence of stars with a core of chirally restored but purely hadronic (confined) matter for masses beyond $1.8~M_\odot$. Stars with pure-quark matter cores are found to be unstable against the gravitational collapse. This instability is shifted to even higher densities if repulsive interactions between quarks are included.
\end{abstract}
\maketitle 

\section{Introduction}
\label{sec:introduction}

 The equation of state (EoS)
  is one of the key observables
  characterizing properties of matter under extreme conditions. In the context of strongly interacting matter, the EoS encodes 
 information on the phase structure and the phase diagram of quantum chromodynamics (QCD).

At a finite temperature and low net-baryon density, the EoS, obtained in ab initio calculations within lattice QCD (LQCD), exhibits a smooth crossover from hadronic matter to a quark-gluon plasma, which is linked to the restoration of the chiral symmetry and color deconfinement  \citep{Aoki:2006we,Bazavov:2018mes}. LQCD provides the state-of-the-art results describing properties of the  low baryon-density  QCD matter that are also successfully  used for the interpretation of data  obtained from  the  relativistic heavy ion collisions
(HIC)~\citep{Ejiri:2005wq,Karsch:2010ck,Borsanyi:2011sw,Bazavov:2012jq,Bazavov:2014pvz,Bazavov:2020bjn}.

The chiral symmetry restoration  is explicitly identified in LQCD calculations  through the scaling properties of the chiral condensate or the behavior of observables that are 
sensitive to chiral criticality, 
such as fluctuations of conserved charges \citep{Bazavov:2020bjn}.  Recently, the chiral symmetry restoration  was explored via the temperature dependence of parity doublers when approaching the QCD phase boundary.
The
LQCD findings~\citep{ Aarts:2015mma,Aarts:2017rrl,Aarts:2018glk} exhibit a clear manifestation of the parity doubling structure for the low-lying baryons around the chiral crossover.
 The masses of the positive-parity ground states are found to be rather temperature-independent, while the masses of negative-parity states drop substantially when approaching  the chiral crossover temperature, $T_c$.   
The parity doublet states become almost degenerate with a finite mass in the vicinity of the chiral crossover. Despite the fact that these LQCD results are still not obtained in the physical limit, 
the observed behavior of parity partners 
 is likely an imprint of the chiral symmetry restoration in the baryonic sector of QCD. 
 It is to be expected that similar results can also appear in the 
cold and  dense nuclear matter when approaching the chiral symmetry restored phase. Indeed, such properties of the chiral partners  
  can be described in 
   the  framework  of 
   the parity doublet model~\citep{Detar:1988kn, Jido:1999hd, Jido:2001nt}. The model has been applied to hot and dense hadronic matter, neutron stars, as well as the vacuum phenomenology of QCD~\citep{Zschiesche:2006zj, Benic:2015pia, Marczenko:2017huu, Marczenko:2018jui, Marczenko:2019trv, Motornenko:2019arp, Mukherjee:2017jzi, Mukherjee:2016nhb, Dexheimer:2012eu, Steinheimer:2011ea, Weyrich:2015hha, Sasaki:2010bp, Yamazaki:2019tuo, Ishikawa:2018yey, Steinheimer:2010ib, Motornenko:2018hjw}.

  Until now,  LQCD calculations at nonzero baryon chemical potential suffer from the sign problem, which stymies their application to the region of low temperature and high net-baryon density. 
  This domain of the QCD phase diagram and the corresponding EoS  are of particular importance for the  understanding  of the extraterrestrial experiments, particularly for the study of compact stellar objects, especially neutron stars (NSs), their mergers~\citep{Bauswein:2018bma}, and supernovae~\citep{Fischer:2017lag}. 
  
  The progress in constraining  the EoS at a low temperature and high density was driven mostly due to discoveries of various two-solar mass NSs~\citep{Demorest:2010bx, Antoniadis:2013pzd, Fonseca:2016tux, Cromartie:2019kug}. The modern observatories for measuring masses and radii of compact objects, the gravitational wave interferometers of the LIGO/Virgo Collaboration (LVC), and the X-ray observatory Neutron star Interior Composition Explorer (NICER) provide new powerful constraints on the neutron-star mass-radius profile. The first ever detection of gravitational waves from the compact star merger GW170817~\citep{Abbott:2018exr}, the second detection from GW190425~\citep{Abbott:2020uma}, as well as the NICER observation of the millisecond pulsar PSR~J0030+451~\citep{Riley:2019yda, Miller:2019cac} delivered simultaneous measurements of masses and radii.
  
  Recent advances in nuclear theory have also tightened the constraints on the EoS over a wide range of densities. This has been achieved by systematic analyses of new astrophysical observations within simplistic approaches, such as the constant-speed-of-sound (CSS) model~\citep{Alford:2013aca} or multipolytropic class of EoSs~\citep{Hebeler:2013nza, Read:2008iy, Alvarez-Castillo:2017qki}. 
  Such schemes are instructive but not microscopic approaches. 
  Although they provide interesting heuristic guidance,  they cannot replace a realistic model for the  EoS.

  One of the  uncertainties when modeling 
  the EoS of dense nuclear matter
  is NS composition. Due to 
  presently limited 
  knowledge, their internal structure remains unclear. Different studies indicate that a variety of exotic constituents may appear in the core of NSs \citep{Blaschke:2016}, such as hyperons or meson condensates (see, e.g.,~\citep{glendenning, Oertel:2016bki}). 
  However, the $2~M_\odot$ constraint is hard to reconcile when additional degrees of freedom are taken into account, since it is known that additional degrees of freedom usually implies   softening of the EoS, and consequently,  
 eventually obtained maximal masses are far below the $2~M_\odot$. This leads, for instance, to the famous hyperon puzzle \citep{Bombaci:2016xzl}.
  One of the most prominent solutions to this problem would be a strong phase transition to deconfined quark matter \citep{Baldo:2003vx,Shahrbaf:2019vtf}.
  Sufficient stiffness of the EoS at high densities is required in order to obtain massive solutions of the Tolman Oppenheimer Volkoff (TOV) equations~\citep{Tolman:1939jz, Oppenheimer:1939ne}. 
  In phenomenological models, it is typically provided by repulsive interactions mediated by the exchange of vector mesons. The inclusion of repulsive vector interactions in quark matter allows for hybrid stars that fulfill the $2~M_\odot$ constraint, as demonstrated in~\cite{Klahn:2006iw}, \cite{Klahn:2013kga}, ~\cite{Benic:2014jia}, and \cite{Klahn:2015mfa}. 
  They are also known to be important for modeling  the QCD phase diagram~\citep{Kitazawa:2003qmg}. 
    At moderate densities, however, the quark matter EoS should be soft enough such that the transition to the deconfined phase  eventually takes place.
    Therefore, a natural question to ask is whether or not massive NSs, such as the recently discovered PSR~J0740+6620 with mass of $2.14^{+0.20}_{-0.18}~M_\odot$~\citep{Cromartie:2019kug}, could contain deconfined quark matter in their core.  This problem has attracted a lot of attention over the past two decades (see, e.g.,~\cite{Trumper:2003we, Ozel:2006km, Alford:2006vz, Klahn:2006iw, Masuda:2012kf, Lenzi:2012xz, Orsaria:2012je, Orsaria:2013hna, Fortin:2020qin}), but no decisive observational evidence has yet been provided. 

  The mechanism of quark confinement and its relation to the chiral symmetry breaking are of major importance in probing the hadron-quark phase transition, although it is nontrivial to embed their interplay into a single effective theory. For this reason, despite its serious shortcomings, the conventional approach is to use separate effective models for the nuclear and quark matter phases (two-phase approaches) with a priori assumed first-order phase transition, typically associated with simultaneous chiral and deconfinement transitions~\citep{Bastian:2015avq}. Major attempts to go beyond the two-phase approach are, for example, the Polyakov loop-extended Nambu--Jona-Lasinio (PNJL)  models~\citep{Fukushima:2003fw, Ratti:2005jh, Sasaki:2006ww, Roessner:2006xn, Fukushima:2008wg}, or the Polyakov loop-extended quark-meson  (PQM)~\citep{Schaefer:2007pw, Skokov:2010wb, Skokov:2010uh, Mizher:2010zb, Herbst:2010rf}. While Polyakov loop-based models provide dynamical means to mimic   confinement of quarks, their caveat is the lack of hadrons as degrees of freedom at low temperature and/or density. Moreover, while at finite temperature and low density the Polyakov loop expectation value serves as an approximated order parameter for  deconfinement, it is highly questionable that it remains so at high density,  which is relevant for astrophysical applications.

  In this work,  we provide the effective dynamical  chiral model  to  quantify the high-density  EoS in the presence of color deconfinement and the chiral symmetry restoration.  
  Instead of using the  Polyakov-loop-based model, we employ the hybrid quark-meson-nucleon (QMN) model~\citep{Benic:2015pia, Marczenko:2017huu, Marczenko:2018jui, Marczenko:2019trv} to explore the implications of dynamical sequential phase transitions at high baryon density on the phenomenology of neutron stars. The model has the characteristic feature that, upon increasing baryon density, the chiral symmetry is restored within the hadronic phase by lifting the mass splitting between chiral partner states, before the quark deconfinement takes place. Quark degrees of freedom are included on top of hadrons, but their unphysical onset is prevented at low densities. This is achieved by an auxiliary scalar field 
  which couples to both nucleons and quarks.
 This field serves as a momentum cutoff in the Fermi-Dirac distribution functions,
 thus it suppresses 
 the unphysical thermal fluctuations of fermions, with the strength linked to the density. 

 The previous implementation 
 of the hybrid QMN model to obtain the nuclear matter  EoS leads to the conclusion,  that the hadronic branch of the mass-radius relation is at tension with the two-solar-mass constraint~\citep{Marczenko:2018jui, Marczenko:2019trv}. In order to improve the description of quark matter properties, we extended the previous hybrid QMN model by including an isoscalar-vector and isovector-vector coupling to quarks. We systematically study the role of the repulsive quark-vector coupling on the deconfinement phase transition and phenomenology of compact stars.

  The hybrid QMN model naturally embeds the interplay between the quark confinement and the chiral symmetry breaking into a single field-theoretical framework, which makes the phenomena inherently connected,
  in contrast 
  to the two-phase approach. These key phenomenological ingredients of the model may have importance not only in the context of modeling of  the interior of NSs, but also in dynamical simulations of their 
  mergers, as well as  supernova collapse and explosions. Previous studies have pointed out the importance of the QCD phase transitions, where the additional latent heat can trigger the local production of neutrinos~\citep{Klahn:2015mfa, Fischer:2016ojn, Klahn:2016uce, Fischer:2017lag, Bauswein:2018bma, Bauswein:2019skm, Fischer:2020xjl, Pereira:2020jgv}. This also motivates further study and possible applications  of the proposed model in the context of HIC at finite temperature, where a novel signature of chiral symmetry restoration within the dense hadronic sector in dilepton production was recently proposed~\citep{Sasaki:2019jyh}.
  
  This paper is organized as follows. In Sect.~\ref{sec:hqmn_model}, we introduce the hybrid quark-meson-nucleon model. In Sect.~\ref{sec:eos}, we discuss the obtained numerical results on the equation of state under neutron-star conditions. In Sect.~\ref{sec:mass_radius}, we discuss the obtained neutron-star relations and confront the results with recent observations. Finally, Sect.~\ref{sec:conclusions} is features our summary and conclusions.



\section{Hybrid QMN model}
\label{sec:hqmn_model}

  \begin{table*}[t!]\begin{center}\begin{tabular}{|c|c|c|c|c|c|}
    \hline
    $m_+~$[MeV] & $m_-~$[MeV] & $m_\pi~$[MeV] & $f_\pi~$[MeV] & $m_\omega~$[MeV] & $m_\rho~$[MeV] \\ \hline\hline
    939   & 1500  & 140     & 93      & 783        & 775 \\ \hline
    \end{tabular}\end{center}
    \caption{Physical vacuum inputs used in this work.}
    \label{tab:vacuum_params}
  \end{table*}

  \begin{table*}[t!]\begin{center}\begin{tabular}{|c|c|c|c|c|c|c|}
    \hline
    $\rho_0~$[fm$^{-3}$] & $E/A - m_+$ [MeV] & $K$~[MeV] & $E_{\rm sym}$~[MeV] & $L~$[MeV] & $m^*_+~[m^{\rm vac}_+]$ & $m^*_-~[m^{\rm vac}_-]$ \\ \hline\hline
    0.16               & -16             & 240       & 31 & 82 & 0.82 & 0.76\\ \hline
    \end{tabular}\end{center}
    \caption{Properties of the nuclear ground state at $\mu_B = 923~$MeV and the symmetry energy used in this work. Also tabulated is the slope of the symmetry energy and the effective masses of the positive- and negative-parity baryonic states at the saturation density given in the units of their vacuum masses.}
    \label{tab:external_params}
  \end{table*}

  \begin{table*}[t!]\begin{center}\begin{tabular}{|c|c|c|c|c|c|c|c|c|c|}
    \hline
    $\lambda_4$ & $\lambda_6f_\pi^2$ & $g^N_\omega$ &  $g^N_\rho$ & $g_1$ & $g_2$ & $g_q$ & $\kappa_b~$[MeV] & $\lambda_b$ & $b_0~$ [MeV]\\ \hline\hline
     33.74          & 13.20           & 7.26       & 7.92 & 13.75 & 7.72 & 3.36& 155 & 0.074 & 569.79\\ \hline
    \end{tabular}\end{center}
    \caption{Numerical values of the model parameters. The values of $\lambda_4$, $\lambda_6,$ and $g^N_\omega$ are fixed by the nuclear ground state properties, $g^N_\rho$ by the symmetry energy, and $g_q$ is fixed by the vacuum quark mass (see the text). The remaining parameters, $\kappa_b$, $\lambda_b,$ and $b_0$ are fixed following \cite{Marczenko:2017huu}.}
    \label{tab:model_params}
  \end{table*}

  In this section, we briefly introduce the hybrid QMN model, which is capable of describing the chiral symmetry restoration and deconfinement phase transitions~\citep{Benic:2015pia, Marczenko:2017huu, Marczenko:2018jui, Marczenko:2019trv}. The hybrid QMN model is composed of the baryonic parity doublet~\citep{Detar:1988kn, Jido:1999hd, Jido:2001nt} and mesons as in the Walecka model~\citep{Walecka:1974qa}, as well as quark degrees of freedom as in the standard linear sigma model~\citep{Scavenius:2000qd}. The spontaneous chiral symmetry breaking yields the mass splitting between the two baryonic parity partners, while it generates the entire mass of a constituent quark. In this work, we consider a system with $N_f=2$; hence, relevant for this study are the positive-parity nucleons, which are proton ($p_+$) and neutron ($n_+$), and their negative-parity partners, denoted as $p_-$ and $n_-$, as well as the up ($u$) and down ($d$) quarks. The fermionic degrees of freedom are coupled to the chiral fields $\left(\sigma, \boldsymbol\pi\right)$, the isosinglet vector-isoscalar field ($\omega_\mu$), and the vector-isovector field ($\boldsymbol \rho_\mu$). The important concept of statistical confinement is realized in the hybrid QMN model by introducing a medium-dependent modification of the particle distribution functions.

  The Lagrangian of the model reads
  \begin{equation}\label{eq:full_lagrangian}
  \mathcal{L} = \mathcal{L}_N +  \mathcal{L}_M + \mathcal{L}_q\textrm,
  \end{equation}
  with $\mathcal{L}_N$, $\mathcal{L}_M$, $\mathcal{L}_q$ denoting the nucleon, meson, and quark parts, respectively. The nucleon part $\mathcal{(L}_N$) of the Lagrangian reads
  \begin{equation}\label{eq:L_n}
  \begin{split}
  \mathcal{L}_N &= i\bar\psi_1\slashchar\partial\psi_1 + i\bar\psi_2\slashchar\partial\psi_2 + m_0\left(  \bar\psi_1\gamma_5\psi_2 - \bar\psi_2\gamma_5\psi_1 \right) \\
  +& g_1\bar\psi_1 \left( \sigma + i\gamma_5 \boldsymbol\tau \cdot \boldsymbol\pi \right)\psi_1 + g_2\bar\psi_2 \left( \sigma - i\gamma_5 \boldsymbol\tau \cdot \boldsymbol\pi \right)\psi_2 \\
  -&g^N_\omega\sum_{k=1,2}\bar\psi_k \slashchar{\omega} \psi_k  -\frac{1}{2}g^N_\rho\sum_{k=1,2}\bar\psi_k \boldsymbol\tau \cdot \slashchar{\boldsymbol\rho} \psi_k \textrm,
  \end{split}
  \end{equation}
  where $\psi_k$ is the set of baryonic chiral fields. Parameters $g_1$, $g_2$, and $g^N_\omega$, $g^N_\rho$ are the baryon-to-meson coupling constants, and $m_0$ is a mass parameter.

  The mesonic part, $\mathcal{L}_M$, of the Lagrangian is introduced as 
  \begin{equation}
  \begin{split}
    \mathcal{L}_M &= \frac{1}{2} \left( \partial_\mu \sigma\right)^2 + \frac{1}{2} \left(\partial_\mu \boldsymbol\pi \right)^2 - \frac{1}{4} \left( \omega_{\mu\nu}\right)^2 - \frac{1}{4} \left( \boldsymbol\rho_{\mu\nu}\right)^2 \\
            &-V_\sigma - V_\omega - V_\rho \textrm,
  \end{split}
  \end{equation}
  where $\omega_{\mu\nu} = \partial_\mu\omega_\nu - \partial_\nu\omega_\mu$ is the field-strength tensor of the vector-isoscalar field, $\boldsymbol\rho_{\mu\nu} = \partial_\mu\boldsymbol\rho_\nu - \partial_\nu\boldsymbol\rho_\mu - g_\rho^N\boldsymbol \rho_\mu \times \boldsymbol \rho _\nu$ is the field-strength tensor of the vector-isovector field, and the potentials read
  \begin{subequations}\label{eq:potentials}
  \begin{align}
    V_\sigma &= -\frac{\lambda_2}{2}\Sigma + \frac{\lambda_4}{4}\Sigma^2 - \frac{\lambda_6}{6}\Sigma^3- \epsilon\sigma \textrm,\label{eq:potentials_sigma}\\
    V_\omega &= -\frac{m_\omega^2 }{2}\omega_\mu\omega^\mu\textrm,\\
    V_\rho &= - \frac{m_\rho^2}{2}{\boldsymbol \rho}_\mu{\boldsymbol \rho}^\mu \textrm,\label{eq:potentials_b}
  \end{align}
  \end{subequations}
  where $\Sigma = \sigma^2 + \boldsymbol\pi^2$, $\lambda_2 = \lambda_4f_\pi^2 - \lambda_6f_\pi^4 - m_\pi^2$, and $\epsilon = m_\pi^2 f_\pi$. $m_\pi$, $m_\omega$, and $m_\rho$ are the $\pi$, $\omega$, and $\rho$ meson masses, respectively, and $f_\pi$ is  the pion decay constant. 
  The parameters $\lambda_4$ and $\lambda_6$ are fixed by the properties of the nuclear ground state. 
  We note that
  the six-point scalar interaction term in Eq.~\eqref{eq:potentials_sigma} is essential in order to reproduce the experimental value of the compressibility \mbox{$K=240\pm20~$MeV~\citep{Shlomo, Motohiro:2015taa}}. Numerical values of  model parameters are  summarized in Tables 
 \ref{tab:vacuum_params},  \ref{tab:external_params}, and \ref{tab:model_params}.

  In the physical basis, the effective masses of the chiral partners, $m_{p_\pm} = m_{n_\pm} \equiv m_\pm$, are given by
  \begin{equation}\label{eq:doublet_masses}
    m_\pm = \frac{1}{2} \left[ \sqrt{\left(g_1+g_2\right)^2\sigma^2+4m_0^2} \mp \left(g_1 - g_2\right)\sigma \right] \textrm.
  \end{equation}
  The positive-parity nucleons are identified as the positively charged and neutral $N(938)$ states: proton ($p_+$) and neutron ($n_+$). Their negative-parity counterparts, denoted as $p_-$ and $n_-,$ are identified as $N(1535)$~\citep{Tanabashi:2018oca}. From Eq.~(\ref{eq:doublet_masses}), it is clear that the chiral symmetry breaking generates only the splitting between the two masses.
  
  When the chiral symmetry is restored, the masses of the baryonic parity partners become degenerate with a common finite mass $m_\pm\left(\sigma=0\right) = m_0$, which reflects the parity doubling structure of the \mbox{low-lying} baryons. Following the previous studies of the \mbox{parity-doublet-based} models~\citep{Zschiesche:2006zj, Benic:2015pia, Marczenko:2017huu, Marczenko:2018jui, Marczenko:2019trv, Motornenko:2019arp, Mukherjee:2017jzi, Mukherjee:2016nhb, Dexheimer:2012eu, Steinheimer:2011ea, Weyrich:2015hha, Sasaki:2010bp, Yamazaki:2019tuo, Ishikawa:2018yey, Steinheimer:2010ib, Motornenko:2018hjw}, as well as recent LQCD results~\citep{Aarts:2017rrl, Aarts:2018glk}, we choose a rather large value: $m_0=700$~MeV. The couplings $g_1$, $g_2$ in Eq.~\eqref{eq:doublet_masses} can be determined by fixing the fermion masses in the vacuum. Their values used in this work are summarized  in Table~\ref{tab:vacuum_params}. Having fixed the chirally invariant mass $m_0$, we obtain the values of the effective masses at saturation density, which are denoted as $m^*_\pm$. We find that the effective mass of the positive-parity states, $m^*_+$, lies within the experimentally determined range~\citep{Kapusta:2006pm, glendenning}. We note, however, that their  exact values depend on the choice of the parameter $m_0$~\citep{Zschiesche:2006zj}. Thus, in principle, the experimental value of $m^*_+$ could be used as an input parameter to constrain the model parameters. In Table~\ref{tab:external_params}, we show the obtained values of the effective masses at saturation density for both parity partners. 

  The quark part$\mathcal{(L}_q$) of the Lagrangian in Eq.~\eqref{eq:full_lagrangian} reads
  \begin{equation}\label{eq:L_q}
  \begin{split}
    \mathcal{L}_q &= \sum_{k=u,d} i\bar\psi_k\slashchar\partial\psi_k + g_q\bar\psi_k \left( \sigma + i\gamma_5 \boldsymbol\tau \cdot \boldsymbol\pi \right)\psi_k  \\
    & -g^q_\omega\sum_{k=u,d}\bar\psi_k \slashchar{\omega} \psi_k -\frac{1}{2}g^q_\rho\sum_{k=u,d}\bar\psi_k \boldsymbol\tau \cdot \slashchar{\boldsymbol\rho} \psi_k \textrm,
  \end{split}
  \end{equation}
  where $\psi_k$ is a set of fields for up (u) and down (d) quarks. Parameters $g^q_\sigma$, $g_\omega^q$ and $g_\rho^q$ are the quark-to-meson coupling constants. The quark effective mass, $m_u = m_d \equiv m_q$, is linked to the sigma field as 
  \begin{equation}\label{eq:mass_quark}
    m_q = g_q \sigma \textrm.
  \end{equation}
  We note that in contrast to the baryonic parity partners (cf. Eq.~\eqref{eq:doublet_masses}), quarks become massless as the chiral symmetry gets restored. The value of the coupling $g_q$ in Eq.~\eqref{eq:mass_quark} can be determined by assuming the quark mass to be $m_q = 1/3~m_+$ in the vacuum.

  The strength of $g^N_\omega$ is fixed by the nuclear saturation properties, while the value of $g^N_\rho$ can be fixed by fitting the value of symmetry energy~\citep{glendenning}. The properties of the nuclear ground state and the symmetry energy are shown in Table~\ref{tab:external_params}. On the other hand, the nature of the repulsive interaction among quarks and their coupling to the $\omega$ and $\rho$ mean fields are still far from consensus. To account for the uncertainty in the theoretical predictions, we treat the couplings $g^q_\omega$ and $g^q_\rho$ as free parameters. 
  We parametrize the couplings of quarks to the $\omega$ and $\rho$ mesons with a single dimensionless parameter $\chi$ as follows:
  \begin{subequations}\label{eq:rep_coupling}
  \begin{align}
    g^q_\omega &= \chi g^N_\omega\textrm,\\
    g^q_\rho &= \chi g^N_\rho\textrm.
  \end{align}
  \end{subequations}

  We stress that the original hybrid QMN model was designed to saturate the Stefan-Boltzmann limit at high temperature and density~\citep{Benic:2015pia, Marczenko:2017huu}. However, the introduction of an additional quark-vector coupling spoils this limit. On the other hand, in this work we considered vector fields only up to quadratic order (cf.~Eq.~\eqref{eq:potentials}), thus, the resulting terms in the effective mean-field potential are also up to quadratic order in densities and do not lead to the acausal behavior of the sound velocity, meaning $c_s^2 \equiv \partial P / \partial \epsilon < 1$ is fulfilled~\citep{Kaltenborn:2017hus}. We note  that it is generally possible to sustain the $2~M_\odot$ constraint and fulfill the conformal bound,$c_s^2 \leq 1/3$. This can be obtained, for example, in a class of constant-speed-of-sound equations of state~\citep{Alford:2015dpa}.

  The hybrid QMN model realizes the concept of {\it \emph{statistical confinement}} through a medium-dependent modification of the Fermi-Dirac distribution functions, where an auxiliary scalar field, $b$ (bag field), is introduced. The distribution functions for the quarks and antiquarks are replaced with
  \begin{subequations}\label{eq:cutoff_quark}
  \begin{align}
          n_q &= \theta \left(\boldsymbol p^2-b^2\right) f_q \textrm,\\
     \bar n_q &= \theta \left(\boldsymbol p^2-b^2\right) \bar f_q \textrm,
  \end{align}
  \end{subequations}
  where $b$ is the expectation value of the $b$-field. The distribution functions for the nucleons and antinuclones are replaced with
  \begin{subequations}\label{eq:cutoff_nuc}
  \begin{align}
           n_\pm &= \theta \left(\alpha^2 b^2 - \boldsymbol p^2\right) f_\pm \textrm,\\
      \bar n_\pm &= \theta \left(\alpha^2 b^2 - \boldsymbol p^2\right) \bar f_\pm \textrm,
  \end{align}
  \end{subequations}
  where $\alpha$ is a dimensionless model parameter,  and $f,\bar f$ are the standard Fermi-Dirac  distribution functions for the particle and antiparticle, respectively. As we demonstrate in Sect.~\ref{sec:eos}, the parameter  $\alpha$ plays a crucial role in tuning the order of the chiral phase transition~\citep{Benic:2015pia, Marczenko:2017huu}.

  From the definition of $n_\pm$ and $n_q$, it is evident that in order to mimic the statistical confinement,
  the  $b$ field should  have a nontrivial vacuum expectation value,  so as 
  to suppress  quark  degrees of freedom    at low densities in the confined phase and to allow for  their population at high densities in the deconfined phase.
  This is achieved by allowing $b$ to be generated from the  potential $V_b$. Following~\cite{Benic:2015pia}, we take the potential of the minimal form
  \begin{equation}\label{eq:bag_potential}
  V_b = -\frac{1}{2}\kappa_b^2 b^2 + \frac{1}{4}\lambda_b b^4 \textrm. 
  \end{equation}
  The potential $V_b$ for a positive $\kappa_b^2$ develops a nontrivial vacuum expectation value at $b_0 = \sqrt{\kappa_b^2 / \lambda_b}$. From Eqs.~(\ref{eq:cutoff_quark}) and (\ref{eq:cutoff_nuc}), one finds that the nucleons favor large $b$, whereas the quarks  favor small $b$. The potential~(Eq. \ref{eq:bag_potential}) is chosen such that, at a certain $T$ and $\mu_B$, a transition sets in, causing the bag-field expectation value to abruptly drop. As a consequence, at low $T$ and $\mu_B$, the quark degrees of freedom are suppressed, while the nucleons get suppressed at high $T$ and $\mu_B$. This characteristic behavior is associated with the deconfinement transition, which is a crucial feature of the model~\citep{Benic:2015pia}.
  
  In Eqs.~\eqref{eq:cutoff_quark} and \eqref{eq:cutoff_nuc}, the functions $f_x$ and $\bar f_x$ are the standard Fermi-Dirac distributions,
  \begin{subequations}
  \begin{align}
    f_x      &= \frac{1}{1+e^{\beta \left(E_x - \mu_x\right)}} \textrm,\\
    \bar f_x &= \frac{1}{1+e^{\beta \left(E_x + \mu_x\right)}}\textrm,
  \end{align}
  \end{subequations}
  with $\beta$ being the inverse temperature and the dispersion relation $E_x = \sqrt{\boldsymbol p^2 + m_x^2}$. The effective chemical potentials for $p_\pm$ and $n_\pm$ are defined as\footnote{In the mean-field approximation, the nonvanishing expectation value of the $\omega$ field is the time-like component; hence we simply denote it by $\omega_0 \equiv \omega$. Similarly, we denote the nonvanishing component of the $\rho$ field, time-like and neutral, by $\rho_{03} \equiv \rho $.}
  \begin{subequations}\label{eq:u_eff_had_iso}
  \begin{align}
    \mu_{p_\pm} &= \mu_B - g^N_\omega\omega - \frac{1}{2}g^N_\rho \rho + \mu_Q\textrm,\\
    \mu_{n_\pm} &= \mu_B - g^N_\omega\omega + \frac{1}{2}g^N_\rho \rho\textrm.
  \end{align}
  \end{subequations}
  The effective chemical potentials for up and down quarks are given by
  \begin{subequations}\label{eq:u_effq}
  \begin{align}
    \mu_u &= \frac{1}{3}\mu_B - g^q_\omega \omega - \frac{1}{2}g^q_\rho \rho + \frac{2}{3}\mu_Q\textrm,\\
    \mu_d &= \frac{1}{3}\mu_B - g^q_\omega \omega + \frac{1}{2}g^q_\rho \rho  - \frac{1}{3}\mu_Q\textrm.
  \end{align}
  \end{subequations}
  In Eqs.~\eqref{eq:u_eff_had_iso}~and~\eqref{eq:u_effq}, $\mu_B$, $\mu_Q$ are the baryon and charge chemical potentials, respectively.

    The concept of momentum cutoff is naturally supported by the asymptotic freedom in QCD, which suggests that active degrees of freedom are different depending on their momenta, meaning hadrons (quarks) with low (high) momenta. Such notion has been widely used in effective theories and the Dyson-Schwinger approach~\citep{Roberts:2010rn,Roberts:2011wy}. In~\citep{Benic:2015pia}, this idea was pushed further by introducing a momentum cut-off as a medium-dependent quantity. Such an intrinsic modification of the cut-off is determined self-consistently when the cut-off is regarded as an expectation value of a scalar field, which is $b$ in our model. As demonstrated in~\cite{Benic:2015pia}, the $b$ field plays a role similar to that of the Polyakov loop at finite temperature and vanishing chemical potential, and it is responsible for the onset of quark degrees of freedom around the crossover temperature. Although the $b$ field is so far not anchored to any QCD symmetry, its introduction is a minimal and practical way to realize the physical situation where quarks are not activated deep in the hadronic phase. Therefore, $b$ is likely related to gluons, especially the chromoelectric component. So far it is not known how one is connected to another, since there is no rigorous order parameter of confinement in QCD with light fermions. Thus, the $b$ field cannot be regarded as a true order parameter. In recent studies, it has been shown that such nontrivial order parameter of the center flavor symmetry exists in a QCD-like theory compactified on a circle~\citep{Cherman:2017tey,Aitken:2017ayq}.
    
    The parameter $\alpha$ manifests itself only in combination with the expectation value of the $b$ field. Therefore, $\alpha$ relates the scales for the momentum distributions of nucleons ($\alpha b$) and quarks ($b$). The lower bound for the $\alpha$ parameter is set so that the ultraviolet cut-off $\alpha b_0$ for nucleon distribution functions and the infrared cut-off $b_0$ for quark distribution functions do not spoil the properties of the nuclear ground state. This sets the lower bound to be roughly $\alpha b_0\geq 300$~\citep{Benic:2015pia, Marczenko:2017huu}. On the other hand, the upper bound on the parameter can be derived by requiring that the system correctly saturates the Stefan-Boltzmann limit, which yields the maximal value for $\alpha b_0 \leq 450~$MeV~\citep{Benic:2015pia, Marczenko:2017huu}. Following our previous works, we chose four representative values within that interval: \mbox{$\alpha b_0 = 350,~370,~400$, and$~450~$MeV}, so as to systematically study the influence of the quark-vector couplings on the equation of state and the resulting mass-radius relations.
    
    \begin{figure*}[t]
  \begin{center}
    \includegraphics[width=0.475\linewidth]{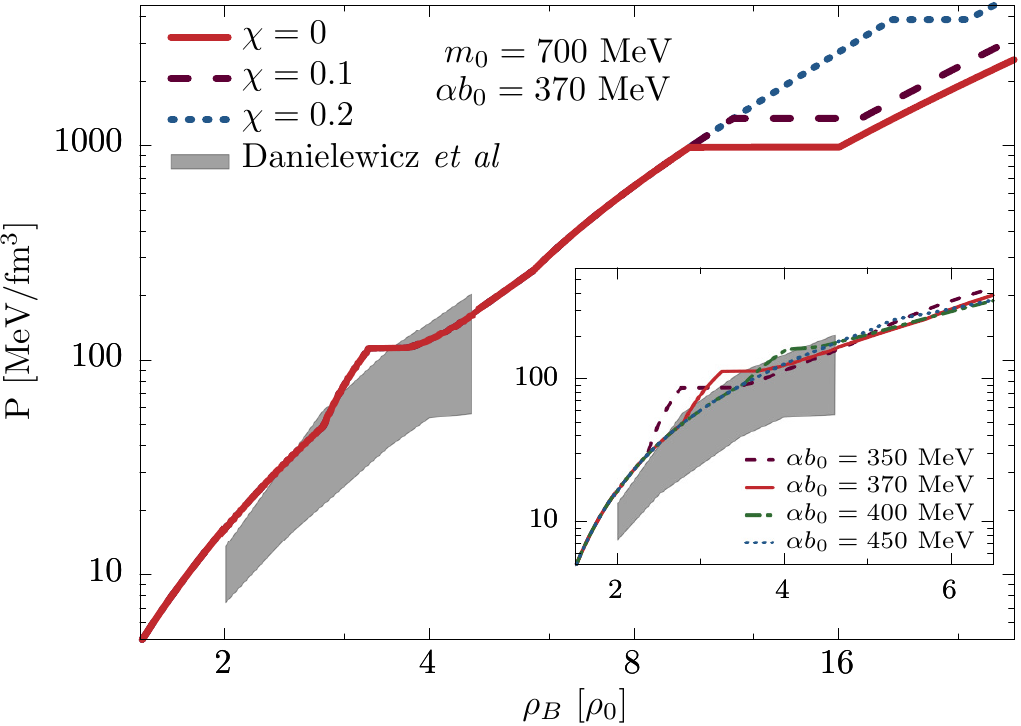}\;\;
    \includegraphics[width=0.475\linewidth]{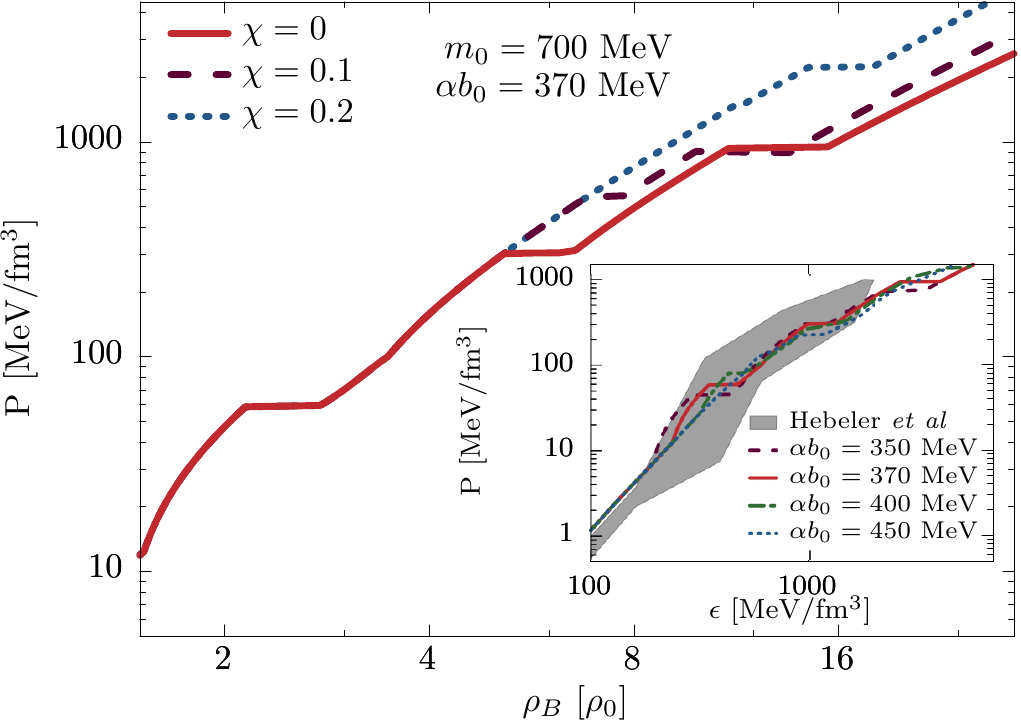}
    \caption{Thermodynamic pressure, $P$, for isospin-symmetric matter (left panel) and under the NS conditions of $\beta$-equilibrium and charge neutrality (right panel), as functions of the net-baryon number density, $\rho_B$, in units of the saturation density, $\rho_0=0.16~$fm$^{-3}$ for $m_0=700~$MeV, \mbox{$\alpha b_0=370~$MeV} and different values of the parameter $\chi$ (see text for details). In the left panel, the flow constraint~\citep{Danielewicz:2002pu} is shown as the gray shaded region. In both panels, the inset plots show pressure for $\chi=0$ and different values of the parameter $\alpha$, in the vicinity of the constraints obtained by~\cite{Danielewicz:2002pu}~(left panel) and~\cite{Hebeler:2013nza} (right panel). We note that the inset figure in the right panel shows pressure as a function of energy density. In both panels, the first-order phase transitions are seen as plateaux of constant pressure.}
    \label{fig:p_n}
  \end{center}
  \end{figure*}
    
     Intuitively, the meaning behind the momentum cut-off $b$ and the $\alpha$ parameter can be understood within a simple bag model picture, where the nucleon is a bound state of three noninteracting massless valence quarks, which are confined in a bag of radius $r$. Due to the uncertainty principle, the quarks cannot have momenta lower than $q_0 \geq 1/r$. With the mass of the ground state nucleon being $m_+ = 3 q_0 = 939~$MeV, this yields the value of the radius $r=0.63~\rm fm$. When the nuclear matter is dense enough, the nucleon wave functions start overlapping and the system exhibits the Pauli blocking effect. In order to estimate the density at which the Pauli blocking manifests itself, one can use the van der Waals excluded volume of hard-sphere nucleons~\citep{Andronic:2012ut}:
    \begin{equation}
        v_{\rm ex} = \frac{16}{3} \pi r^3 \textrm,
    \end{equation}
    which corresponds to the closest packing density of $\rho_c = 1/v_{\rm ex} = 0.238~{\rm{fm}}^{-3} = 1.49 \rho_0$, where $\rho_0=0.16~\rm fm^{-3}$ is the saturation density. The density $\rho_c$ naturally corresponds to the onset of the stiffening effect in the hybrid QMN model. At higher density, the wave functions of nucleons overlap and the quark momenta have to be rearranged, in order not to violate the Pauli exclusion principle. A simple argument how the Pauli principle may act in a dense gas of nucleons as three-quark bound states is that the higher quark momenta have to be populated. At zero temperature, the shell in the quark-momentum space, $\Delta \rho_q$, which has to be filled, ranges from $q_0$ to $q_0+\delta q_0$ and corresponds to the quark density
    \begin{equation}
    \Delta \rho_q = 3 \Delta \rho_B = 3\gamma_q\int\limits_{q_0}^{q_0+\delta q_0} \mathrm{d} q \; \frac{q^2}{2 \pi^2} =\frac{\gamma_q}{2\pi^2} q_0^2 \delta q_0 + \mathcal{O}(\delta q_0^2) \textrm,
    \end{equation}
    where $\gamma_q$ is the spin-color-flavor degeneracy factor of quarks. From the above, the shift of the quark momentum, $\delta q_0$, can be related to the corresponding Pauli shift in the nucleon energy, 
    $\Delta^{\rm Pauli} = 3\delta q_0 = 18 \pi^2/(\gamma_q q_0^2) \Delta\rho_B$. At densities exceeding the critical density, $\rho_c$, where the effect sets in, the Pauli shift corresponds to a density-dependent lowering of the chemical potential. The critical density can be translated to the critical Fermi momentum of nucleons,
    \begin{equation}
    p_f = \sqrt[3]{\frac{3}{2} \pi^2 \rho_c} = \sqrt[3]{\frac{9}{32} \pi} \; q_0 \textrm,
    \end{equation}
    and compared with the corresponding UV cut-off $\alpha b_0$ in the hybrid QMN model. The value $q_0$ is identified with the IR cut-off in the quark momentum distribution due to confinement and the uncertainty relation. In the hybrid QMN model, it is given by $b_0$. Thus, one gets the UV cut-off in the nucleon distribution to be $p_f = \alpha b_0$. From this, one gets $\alpha = \sqrt[3]{9\pi/32} \approx 0.959$. We note that the maximal value of the $\alpha$ parameter obtained in the hybrid QMN model is $\alpha_{\rm max} = 2^{-1/3} \approx 0.794$~\citep{Benic:2015pia,Marczenko:2017huu}. This simple analysis serves as a ballpark estimate for the parameter $\alpha$ and shows that it may be quantified by considering the Pauli exclusion principle on the quark level. The present estimate within an intuitive bag model picture and the van der Waals excluded volume leads to similar scaling as the one obtained in the hybrid QMN model, meaning smaller values of $b_0$ and $q_0$ lead to an earlier onset of the stiffening effect. A more realistic calculation of the quark Pauli blocking in nuclear matter can be found in~\citep{Blaschke:2020qrs}.
  
  The thermodynamic potential in the mean-field approximation reads~\citep{Marczenko:2018jui}
  \begin{equation}\label{eq:thermo_pot_iso}
     \Omega = \sum_{x=p_\pm,n_\pm,u,d}\Omega_x + V_\sigma + V_\omega + V_\rho + V_b \textrm,\end{equation}
  where the summation goes over the fermionic degrees of freedom. The kinetic part,  $\Omega_x$,  of the thermodynamic potential  Eq.~\eqref{eq:thermo_pot_iso}, reads
  \begin{equation}\label{eq:thermokin}
     \Omega_x = \gamma_x \int\frac{\dd^3p}{\left(2\pi\right)^3} T \left[\ln\left(1-n_x\right) + \ln\left(1-\bar n_x\right)\right]\textrm,
  \end{equation}
  where the functions $n_x$ and $\bar n_x$ are the modified Fermi-Dirac distributions defined in Eqs.~\eqref{eq:cutoff_quark} and \eqref{eq:cutoff_nuc}. The spin degeneracy factor, $\gamma_x$ for nucleons is $\gamma_\pm=2$ for both positive- and negative-parity states, while the spin-color degeneracy factor for quarks is $\gamma_q=2\times 3 = 6$.

  The physical inputs and the model parameters used in this work are summarized in Tables~\ref{tab:vacuum_params} and~\ref{tab:external_params}. In-medium profiles of the mean fields are obtained by extremizing the thermodynamic potential~in Eq.~\eqref{eq:thermo_pot_iso}, leading to the following gap equations:
  
  \begin{subequations}\label{eq:gap_eqs_iso}
  \begin{align}
    \frac{\partial\Omega}{\partial\sigma} &= \frac{\partial V_\sigma}{\partial \sigma} + \sum_{x=p_\pm,n_\pm,u,d}s_x \frac{\partial m_x}{\partial \sigma}  \textrm, \label{eq:gap_eq_sigma}\\
    \frac{\partial\Omega}{\partial\omega} &= \frac{\partial V_\omega}{\partial \omega} + \sum_{x=p_\pm,n_\pm,u,d}g^x_\omega\rho_x  \textrm,\label{eq:gap_omega}\\
    \frac{\partial\Omega}{\partial\rho}   &= \frac{\partial V_\rho}{\partial \rho} + \sum_{x=p_\pm, u}\frac{g^x_\rho}{2}\rho_x - \sum_{x=n_\pm, d}\frac{g^x_\rho}{2}\rho_x  \textrm,\\
    \frac{\partial\Omega}{\partial b}     &= \frac{\partial V_b}{\partial b} + \alpha \sum_{x=p_\pm,n_\pm} \omega_x - \sum_{x=u,d}\omega_x  \textrm,\label{eq:gap_b}
  \end{align}
  \end{subequations}
  where the scalar and baryon densities are
 \begin{equation}\label{eq:scalar_den}
      s_x = \gamma_x \int\frac{\dd^3 p}{(2\pi)^3}\; \frac{m_x}{E_x} \left( n_x + \bar n_x \right) \end{equation}
    and
  \begin{equation}\label{eq:vector_den}
      \rho_x = \gamma_x \int\frac{\dd^3 p}{(2\pi)^3}\; \left( n_x - \bar n_x \right) \textrm,
    \end{equation}
    respectively. The terms $\omega_x$ in the gap Eq.~\eqref{eq:gap_b} are given as
  \begin{equation}\label{eq:boundary_nucleon}\small
      \omega_{\pm} = \gamma_{\pm} \frac{(\alpha b)^2}{2\pi^2}T\left[ \ln\left(1 - f_{\pm}\right) + \ln\left(1 - \bar f_{\pm}\right) \right]_{\boldsymbol p^2 = (\alpha b)^2}
\end{equation}
    and
  \begin{equation}\label{eq:boundary_quark}\small
    \omega_q = \gamma_q \frac{b^2}{2\pi^2} T \left[ \ln\left(1 - f_q\right) + \ln\left(1 - \bar f_q\right) \right]_{\boldsymbol p^2 = b^2} \
\end{equation}
  for the nucleons and quarks, respectively. We note that the terms in Eqs.~\eqref{eq:boundary_nucleon} and \eqref{eq:boundary_quark} come into the gap Eq.~\eqref{eq:gap_b} with opposite signs. This reflects the fact that nucleons and quarks favor different values of the bag~field. 

  In the grand canonical ensemble, the~thermodynamic pressure is obtained from the thermodynamic potential as \mbox{$P = -\Omega + \Omega_0$}, where $\Omega_0$ is the value of the thermodynamic potential in the vacuum. The~net-baryon number density for a species $x$ is defined as
  \begin{equation}
    \rho^x_B = -\frac{\partial \Omega_x}{\partial \mu_B} \textrm,
  \end{equation}
  where $\Omega_x$ is the kinetic term in Eq.~\eqref{eq:thermokin}. The~total net-baryon number density~reads
  \begin{equation}
    \rho_B = \rho_B^{n_+} + \rho_B^{n_-} + \rho_B^{p_+} + \rho_B^{p_-} + \rho_B^{u} + \rho_B^{d} \textrm.
  \end{equation}
  The particle-density fractions are defined as
  \begin{equation}\label{eq:fractions}
    Y_x = \frac{\rho_B^x}{\rho_B} \textrm.
  \end{equation}
  The symmetry energy and its slope at saturation density are given as
 \begin{equation}\label{eq:e_sym}
      E_{\rm sym} = \frac{1}{2}\frac{\partial^2 \left(\epsilon / \rho_B \right)}{\partial \delta^2}\Bigg|_{\delta = 0} 
  \end{equation}
  and
  \begin{equation}
      L = 3 \rho_0 \frac{\partial E_{\rm sym}}{\partial \rho_B} \Bigg|_{\rho_B = \rho_0}\textrm,
  \end{equation}
  respectively. In Eq.~\eqref{eq:e_sym}, $\epsilon$ is the energy density and the asymmetry parameter \mbox{$\delta = (\rho^{n^+}_B + \rho^{n^-}_B - \rho^{p^+}_B - \rho^{p^-}_B + \rho_B^d - \rho_B^u)/\rho_B$}\footnote{We note that this generalized definition reduces to the known definition of the asymmetry parameter, \mbox{$\delta = (\rho_B^{n^+} - \rho_B^{p^+}) / \rho_B$}, where the negative parity states and quarks are not populated and their densities vanish, which is the case at the saturation density.}. The value of the slope of the symmetry energy at saturation density obtained in the model is $L=82~$MeV. The value is rather large when compared to the recent chiral effective field theory estimate~\citep{Kruger:2013kua}. We note that similar values are found in other parity-doublet models at the mean-field-level thermodynamics~\citep{Motohiro:2015taa}, and it is consistent with the commonly considered range for the parameter~\citep{Oertel:2016bki, Li:2013ola}.

  In the following section, the above hybrid QMN model equation of state  of strongly interacting matter will be applied  to identify properties of compact stellar objects such as NSs. 
   
\section{Equation of state}
\label{sec:eos}

\begin{figure}[t]
  \begin{center}
    \includegraphics[width=.95\linewidth]{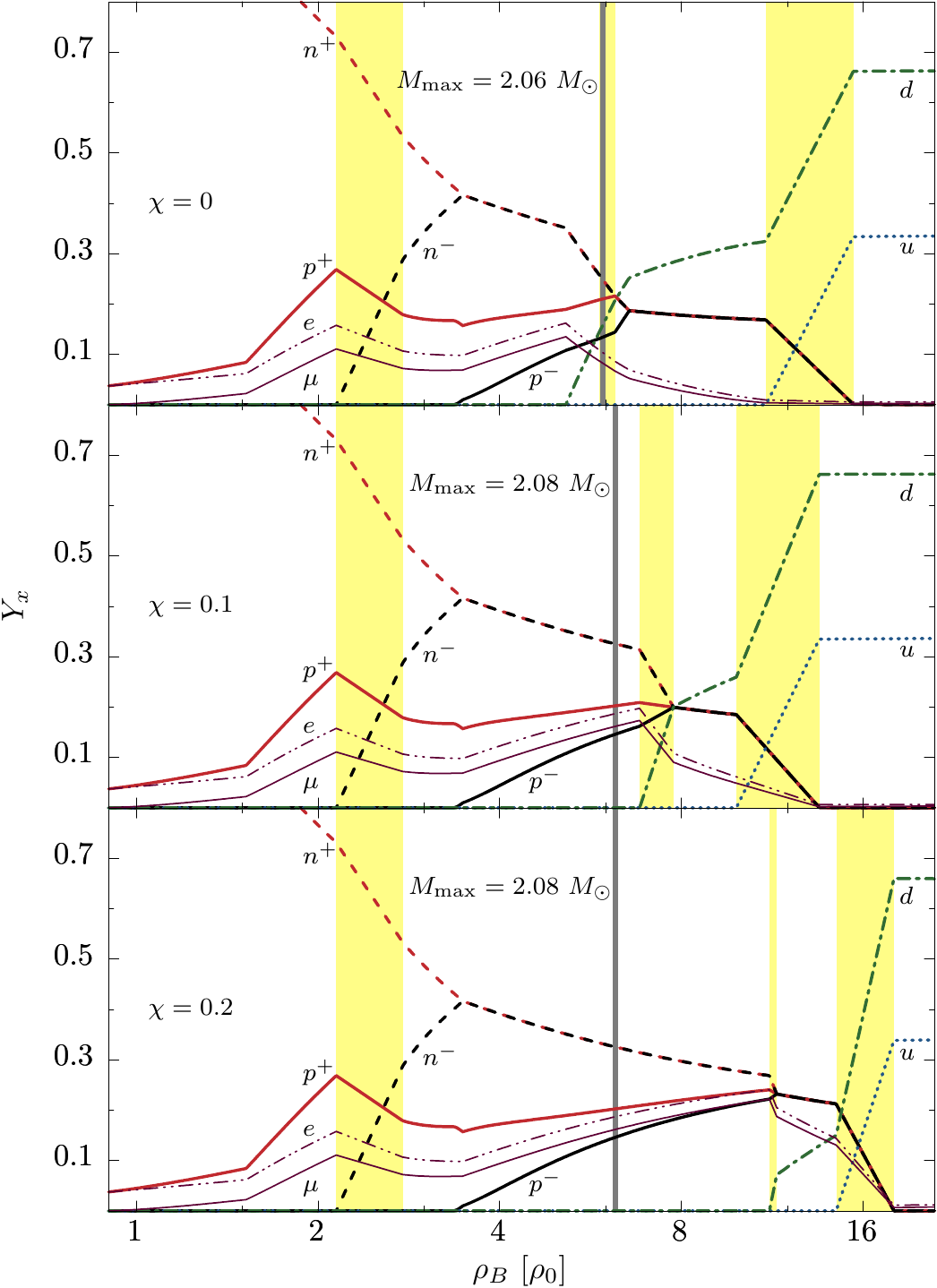}
    \caption{Composition of compact star matter as a function of the central baryon density  in the units of saturation density for the present model for $m_0 = 700~$MeV and $\alpha b_0 = 370~$MeV. Three different cases are considered for \mbox{$\chi=0$} (top panel), \mbox{$\chi=0.1$} (middle panel), \mbox{$\chi=0.2$} (bottom panel). In each panel, the yellow shaded areas indicate the density regions of phase coexistence. The gray, solid vertical lines indicate the central densities of the associated maximum-mass neutron stars.}
    \label{fig:composition}
  \end{center}
  \end{figure}

  We started by generalizing the hybrid QMN EoS derived for the cold nuclear matter to account for the isospin effects,  considering  isospin-symmetric 
  and  highly asymmetric environments such as the interiors of NSs. 
 The composition of neutron-star matter requires \mbox{$\beta$ equilibrium} with electrons~($e$) and muons~($\mu$), included as free relativistic particles, as well as the charge neutrality condition. To this end, we solve the set of equations,
  \begin{subequations}
  \begin{align}
    &\mu_e = \mu_\mu = \mu_{n_+} + \mu_{n_-} - \mu_{p_+} - \mu_{p_-} + \mu_d - \mu_u  \\
    &\rho_{p_+} + \rho_{p_-} + \frac{2}{3}\rho_u -\frac{1}{3}\rho_d - \rho_e - \rho_\mu = 0\textrm,
  \end{align}
  \end{subequations}
  at $T=0$, where $\mu_x$'s and $\rho_x$'s are the effective chemical potentials of given species (see Eqs.~\eqref{eq:u_eff_had_iso} and \eqref{eq:u_effq}) and their densities (see Eq.~\eqref{eq:vector_den}), respectively.
  
  \begin{table*}[t!]\begin{center}\begin{tabular}{|c||c|c|c|c|}
    \hline
                & \multicolumn{4}{c|}{$\alpha b_0~$ [MeV]}         \\ \hline
    $\chi$      & $350$                    & $370$           & $400$           & $450$    \\ \hline\hline
    \multirow{3}{*}{0}   & $1.82 - 2.60$   & $2.14 - 2.76$   & $2.61 - 2.92$   & $3.56$ \\
                         & $4.98 - 6.11$   & $5.84 - 6.21$   & $5.10$          & $4.82 - 6.04$ \\
                         & $9.21 - 13.58$  & $11.03 - 15.42$ & $16.40 - 19.25$ & $10.84$ \\ \hline
    \multirow{3}{*}{0.1} & $1.82 - 2.60$   & $2.14 - 2.76$   & $2.61 - 2.92$   & $3.56$  \\
                         & $6.56 - 7.20$   & $6.82 -7.75$    & $7.13 - 8.32$   & $6.61$  \\ 
                         & $8.48 - 12.32$  & $9.86 - 13.54$  & $11.96 - 15.89$ & $15.67-19.63$  \\ \hline
    \multirow{3}{*}{0.2} & $1.82 - 2.60$   & $2.14 - 2.76$   & $2.61 - 2.92$   & $3.56$  \\ 
                         & $12.07 - 15.67$ & $11.19 - 11.51$ & $11.97 - 12.59$ & $11.89 - 12.82$   \\ 
                         & $12.07 - 15.67$ & $14.38 - 18.00$ & $17.89 - 21.72$ & $23.28 - 27.56$ \\ \hline
    \end{tabular}\end{center}
    \caption{Baryon density ranges of the coexistence phases associated with the chiral restoration (top), onset of down (middle) and up (bottom) quark under the neutron-star conditions, in terms of saturation density units, $\rho_0$, for different values of $\alpha b_0$ and $\chi$ parameters. In the cases where transitions proceed as smooth crossovers, a single value is given.}
    \label{tab:transitions_dens}
  \end{table*}

  In Fig.~\ref{fig:p_n}, we show the calculated zero-temperature equations of state in the mean-field approximation for selected parametrization, $m_0=700~$MeV and $\alpha b_0=370~$MeV, for different strengths of the quark vector-interaction $\chi$ (see Eq.~\eqref{eq:rep_coupling}), namely \mbox{$\chi=0$} (red, solid line), \mbox{$\chi=0.1$} (purple, dashed line), and \mbox{$\chi=0.2$} (blue, dotted line), as functions of the baryon density in the units of saturation density. In the left panel of the figure, we show the isospin-symmetric matter EoSs. In each case, the behavior at low density is similar. Shown EoSs feature a common first-order chiral phase transition, determined as a jump in the $\sigma$-field expectation value, which causes the parity-doublet nucleons to become almost degenerate with the mass $m_\pm = m_0$. The chiral phase transition is triggered at roughly $3.25~\rho_0$, with the mixed phase persisting up to $3.68~\rho_0$. The mechanism of statistical confinement has a prominent twofold impact on the class of EoSs obtained in the model. First, the strength of the chiral phase transition depends on the choice of $\alpha$. Namely, higher values yield weaker first-order transition, which turns into a second-order and eventually becomes a smooth crossover, defined as a peak in $\partial\sigma/\partial \mu_B$. Second, lower values of $\alpha$ result in a stiffening of the EoS, due to which, the chiral phase transition is triggered at lower densities.~\citep{Marczenko:2017huu}. We illustrate the stiffening of the EoS based on a simplified model in Appendix~\ref{sec:appendixA}.

  We note  that the stiffening effect of nuclear matter just above the saturation density is a feature that the present approach has in common with, for example, the relativistic density-functional models of nuclear matter with excluded nucleon volume (such as the one by \cite{Typel:2016srf}). These are used, for example, in the effective multi-polytrope \citep{Alvarez-Castillo:2017qki} and CSS~\citep{Alford:2017qgh} class of hybrid EoS. However, too extreme stiffening could represent a certain tension with the recent analysis of GW170817~\citep{Abbott:2018exr}. In principle, this tension could be resolved, for example, by a strong phase transition occurring in the compact star mass range relevant for GW170817 (see reference~\citep{Paschalidis:2017qmb}).

  In the left panel of Fig.~\ref{fig:p_n}, we also show the proton flow constraint (gray region) obtained in heavy ion collisions~\citep{Danielewicz:2002pu}, where the matter is assumed to be isospin-symmetric. The agreement with the EoSs obtained in the hybrid QMN model is fairly good. Shown EoSs go through the upper band of the constraint. Interestingly, the chiral phase transition happens within the baryon-density range of the flow constraint. The inset plot in the left panel of Fig.~\ref{fig:p_n} shows the EoSs obtained for different values of the $\alpha$ parameter in the vicinity of the flow constraint. Equations of state for $\alpha b_0=350,~370$, and$~400~$MeV feature the first-order chiral phase transition within this constraint. The softest EoS for $\alpha b_0 = 450~$MeV features a smooth crossover transition slightly above the densities governed by the flow constraint, at roughly $5.4~\rho_0$. The stiffening of the EoSs is evidently seen before the chiral phase transition takes place. This highlights the importance of the chiral phase transition for the density range accessible in heavy ion collisions. We note  that the  chiral symmetry restoration  
  in the dense hadronic sector can possibly be identified in the  production rate of dileptons,  
  as proposed recently in~\cite{Sasaki:2019jyh}. 

  As the density increases, the equation of state with no quark-vector coupling, that is, \mbox{$\chi=0$}, features another jump in baryon density, associated with the deconfinement of quarks, with a mixed phase from $9.65 - 16.04~\rho_0$. When the inclusion of the quark-vector coupling is considered, meaning the \mbox{$\chi=0.1$} and \mbox{$\chi=0.2$} cases in Fig.~\ref{fig:p_n}, the onset of both quarks is systematically shifted toward higher densities, when compared to the case with vanishing coupling, with mixed phases between $11.19-17.28~\rho_0$ and $19.18-24.83~\rho_0$, respectively. Consequently, the hadronic part of the EoS is extended beyond the deconfinement in the case of vanishing $\chi$. We note that all three equations of state remain the same up to the point where the quarks start to  appear. We stress that the chiral and hadron-to-quark phase transitions are sequential.

  In the right panel of Fig.~\ref{fig:p_n}, we show the corresponding EoSs under the neutron-star conditions for $m_0=700~$MeV and $\alpha b_0=370~$MeV. Similarly to the isospin-symmetric case, the behavior at low densities is quite comparable. Shown EoSs feature a common first-order chiral phase transition. For comparison, the EoSs obtained for the remaining values of the parameter $\alpha$ are also shown in the vicinity of the chiral phase transitions. We note  that the chiral phase transitions occur at systematically lower densities in comparison to the isospin-symmetric matter. However, as the density increases, the EoSs feature another two sequential jumps in baryon density. The first is associated with the onset of the down quark, and the second is associated with the onset of the up quark, after which the EoS is composed solely of quarks. The sequential appearance of quarks in the isospin-asymmetric matter stays in contrast to the case of symmetric case, where quarks deconfine simultaneously, owing to the isospin symmetry. This can be traced back to their different electric charge, as well as to the coupling to the $\rho$ meson, which introduces different signs for both quarks in their effective chemical potentials (see Eq.~\eqref{eq:u_effq}). Thus, the obtained EoSs under the neutron-star conditions feature sequential (i.e., flavor-dependent) deconfinement phase transition. We stress that the sequential deconfinement of quarks under the neutron-star conditions cannot be obtained in a class of pure quark models or even in a class of hadron-quark models, where the confined and deconfined phases are treated independently. This is due to the negative charge of the down quark, which cannot be neutralized with negatively-charged leptons. The effect of the finite quark-vector coupling is similar to the isospin-symmetric case. Namely, the onset of both quarks is systematically shifted toward higher densities when compared to the case with vanishing coupling. Consequent extension of the hadronic branch of the EoS is exhibited. In this case, the EoSs remain the same up to the point where the down quark appears. This behavior is also resembled by the density fractions shown in Fig.~\ref{fig:composition}. The composition of the compact-star matter for $m_0 = 700~$MeV and $\alpha b_0 = 370~$MeV and \mbox{$\chi=0$} (top panel), \mbox{$\chi=0.1$} (middle panel), \mbox{$\chi=0.2$} (bottom panel), in terms of the particle-density fractions, $Y_x$, defined in Eq.~\eqref{eq:fractions}, as a function of the baryon density in the units of saturation density is also illustrated in  Fig.~\ref{fig:composition}. Similarly to the EoS, the compositions at low density are the same, up to the point of the appearance of the down quark in the case with \mbox{$\chi=0$} (green, dash-dotted line in the top panel). With a finite value of the parameter $\chi$, the hadronic part of the EoS is extended, and consequently the deconfinement of quarks is shifted to higher densities. Because the down quark always appears before the up quark, due to lower Fermi momenta, in order to fulfill the charge neutrality constraint, its negative electric charge can only be neutralized by the positively-charged hadrons, meaning protons, $p_+$, and its chiral partners, $p_-$. Such a mixed phase, comprised of hadrons and the down quark, persists until the hadrons become suppressed and the up quark is populated, compensating for the positively-charged hadrons in order to meet the charge neutrality condition. We note that in the case of isospin-asymmetric matter, the phase transitions are triggered at lower densities. The baryon-density jumps associated with consecutive sequential transitions for the sets of model parameters used in this work are shown in Tab.~\ref{tab:transitions_dens}.

  \begin{figure}[t]
  \begin{center}
    \includegraphics[width=0.95\linewidth]{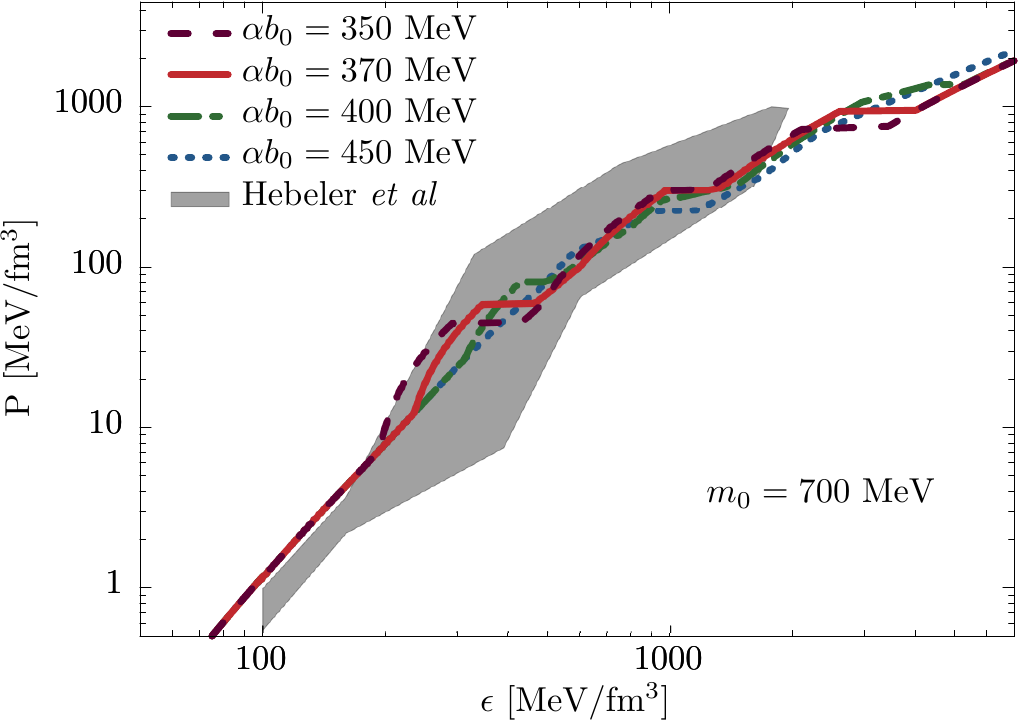}
    \caption{Thermodynamic pressure, $P$, under the NS conditions of $\beta$-equilibrium and charge neutrality, as a function of the energy density, $\epsilon$, for $m_0 = 700~$MeV and $\chi = 0$, and four representative values of the parameter $\alpha$ (see text for details). The phase transitions are seen as plateaux of constant pressure. The gray shaded region marks the constraint obtained by~\cite{Hebeler:2013nza}.}
    \label{fig:p_e}
  \end{center}
  \end{figure}
  
\begin{figure}[t]
  \begin{center}
    \includegraphics[width=0.95\linewidth]{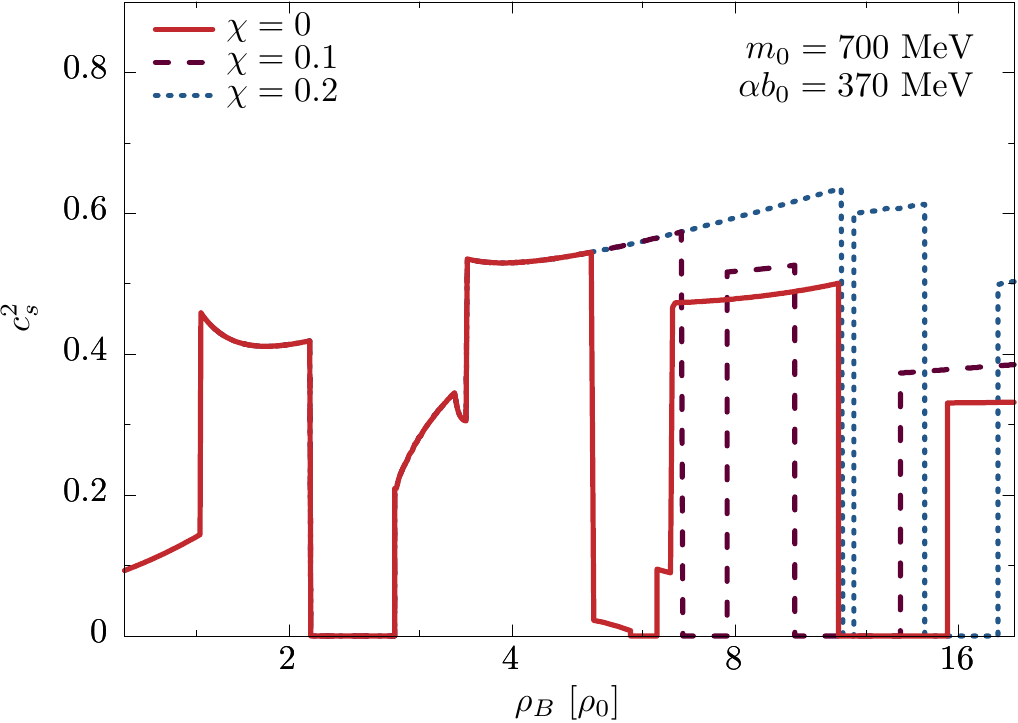}
    \caption{Speed of sound squared as a function of the net-baryon number density, $\rho_B$, in units of the saturation density, $\rho_0=0.16 \rm fm ^{-3}$ for $m_0=700~$MeV, $\alpha b_0=370~$MeV and different values of the parameter $\chi$ (see text for details). The first order phase transitions are seen as plateaux of vanishing speed of sound.}
    \label{fig:cs2}
  \end{center}
  \end{figure}

  In Fig.~\ref{fig:cs2}, we show the speed of sound squared, \mbox{$c_s^2 = \dd P / \dd \epsilon$}, in units of the speed of light squared, as a function of the net-baryon number density. The coexistence phases due to first-order phase transitions are seen as plateaux of vanishing speed of sound. The notable swift increases of the speed of sound in each curve are results of the stiffening mechanism that arises due to the statistical confinement implemented in the model (cf. Eq.~\eqref{eq:cutoff_nuc}).

  We note that there is a theoretical uncertainty in the prediction for the ratio of the vector- and scalar-quark couplings in the class of Nambu--Jona-Lasinio (NJL) and the Polyakov loop-extended NJL models~\citep{Sasaki:2006ws,Fukushima:2008wg, Bratovic:2012qs, Orsaria:2012je, Orsaria:2013hna, Klahn:2013kga}. Interestingly, we find that the  results are essentially  independent of  the exact value of the scalar coupling, $g_q$ in the hybrid QMN model. This is because the value of the coupling is fixed to the the properties in the vacuum, where quarks are not expected to be present as effective degrees of freedom. In the hybrid QMN model, this is controlled  by the finite expectation value of the $b$-field, which suppresses the quark contribution to the gap equations at low density (see Sect.~\ref{sec:hqmn_model}). On the other hand, the quarks are populated at high densities. There, however, the expectation value of the scalar mean field, $\sigma$, is of the order of a few MeV, and slowly vanishes as the density increases. Therefore, the quark contribution to the gap equation in the $\sigma$ direction is negligible. Thus, contrary  to other effective models, the laxity in the value of the coupling, $g_q,$ is a direct consequence of the confinement mechanism implemented in the model, and seems natural from the phenomenological point of view. We note that at high temperatures it might not persist, owing to additional thermal fluctuations.
  
  We stress that all  EoSs discussed in this work fall into the region derived from the maximum mass constraint $M_{\rm max}\geq 2.01(4)~M_\odot$~\citep{Antoniadis:2013pzd} obtained by Hebeler {\it et al} in Ref.~\cite{Hebeler:2013nza} using a multi-polytrope ansatz for the EoS above the saturation density. This is shown in Fig.~\ref{fig:p_e} for the case of vanishing parameter $\chi$. Interestingly, the chiral phase transition and the deconfinement of down quark lie within the region set by the constraint. This is in contrast to the symmetric-matter flow constraint, where only the chiral transition is featured within the constraint. The up quarks are populated only at higher values of the energy density. We note that  for high enough value of the parameter $\chi$, the deconfinement transition would shift to higher density and eventually fall out of the constraint.

  In general, the class of EoSs obtained in the hybrid QMN model for the case of isospin-asymmetric matter consist of four consecutive density regions, from low to high density: confined and chirally broken phase, confined and chirally restored phase, partially deconfined, and fully deconfined phase. The intermediate phases are for the novel states in which confined and deconfined matter may coexist. Such remarkable separation of the chirally broken and the deconfined phase might indicate the existence of a quarkyonic phase, where the quarks are partly confined to form a Fermi surface, but the relevant degrees of freedom remain the nucleons with the restored chiral symmetry~\citep{Hidaka:2008yy, McLerran:2008ua, Andronic:2009gj, McLerran:2008ua, McLerran:2018hbz, Jeong:2019lhv}. Furthermore, in Ref.~\cite{Blaschke:2008br}, it was conjectured that a mixed phase, comprised of nucleons and down  quarks, may have compelling nontrivial cooling and transport properties that are of interest for the thermal and rotational evolution of compact stars.

  \begin{figure*}[t]
  \begin{center}
    \includegraphics[width=0.475\linewidth]{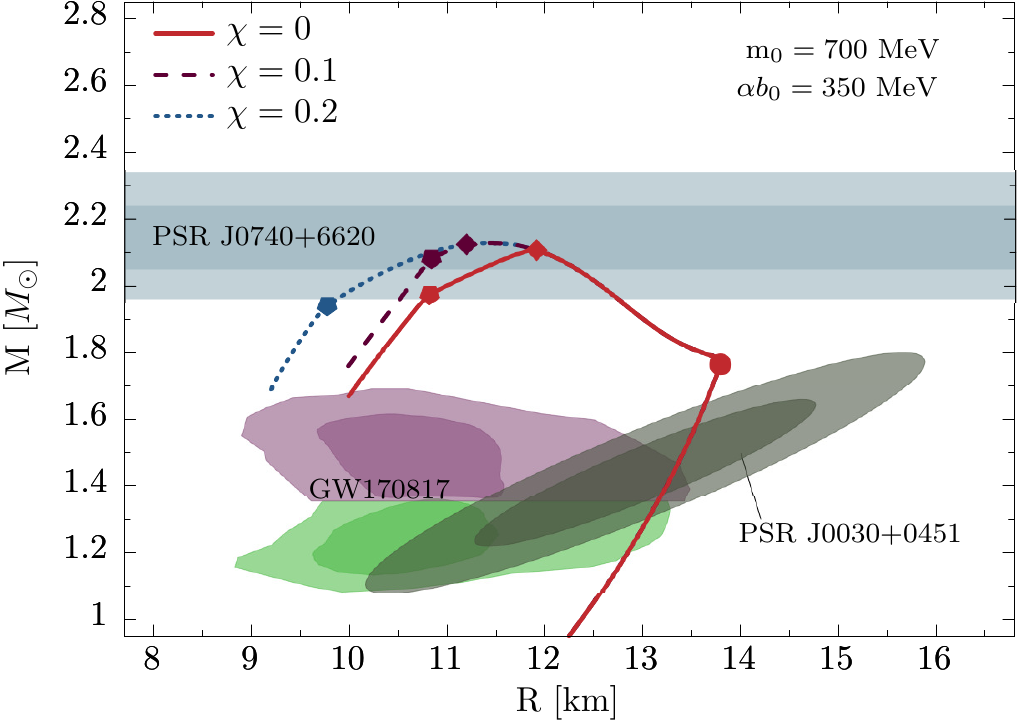}\;\;
    \includegraphics[width=0.475\linewidth]{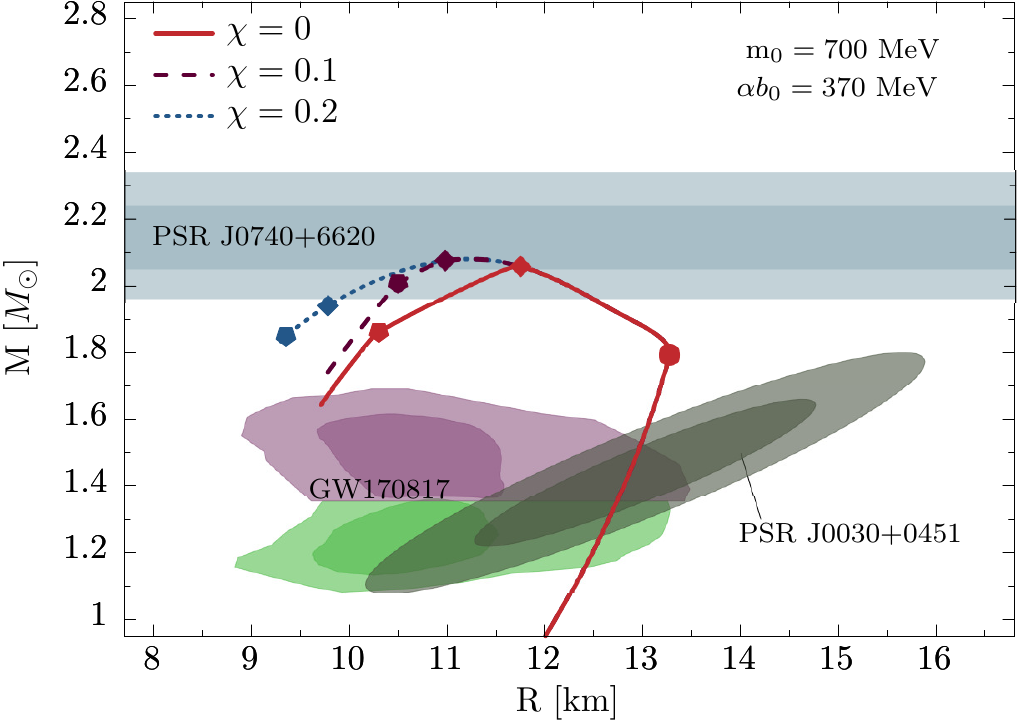}

    \includegraphics[width=0.475\linewidth]{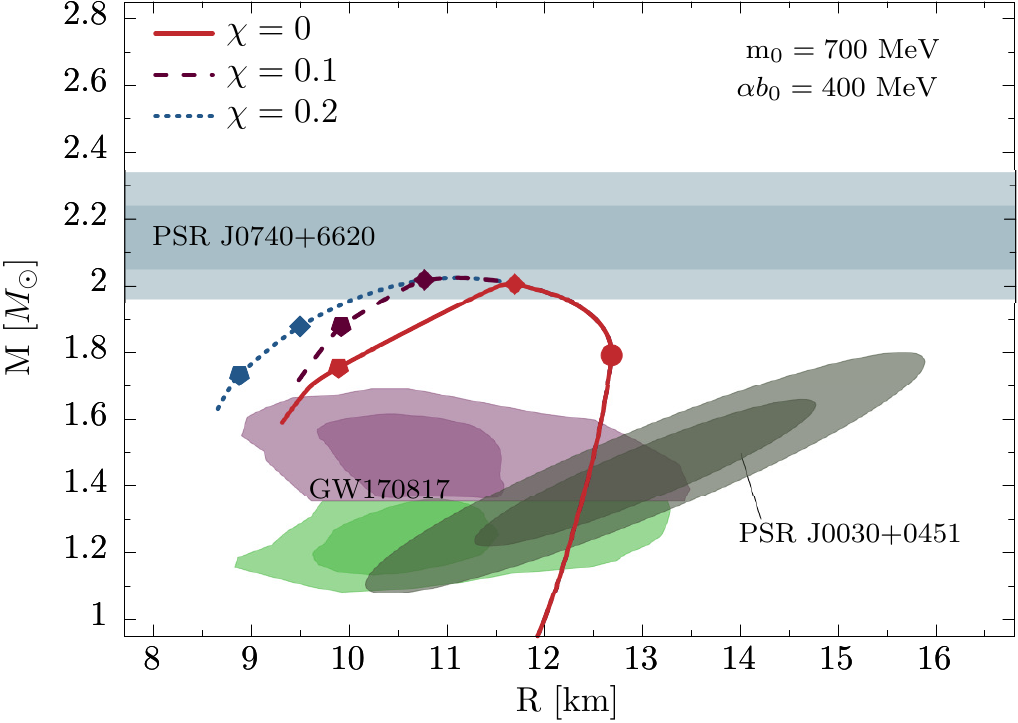}\;\;
    \includegraphics[width=0.475\linewidth]{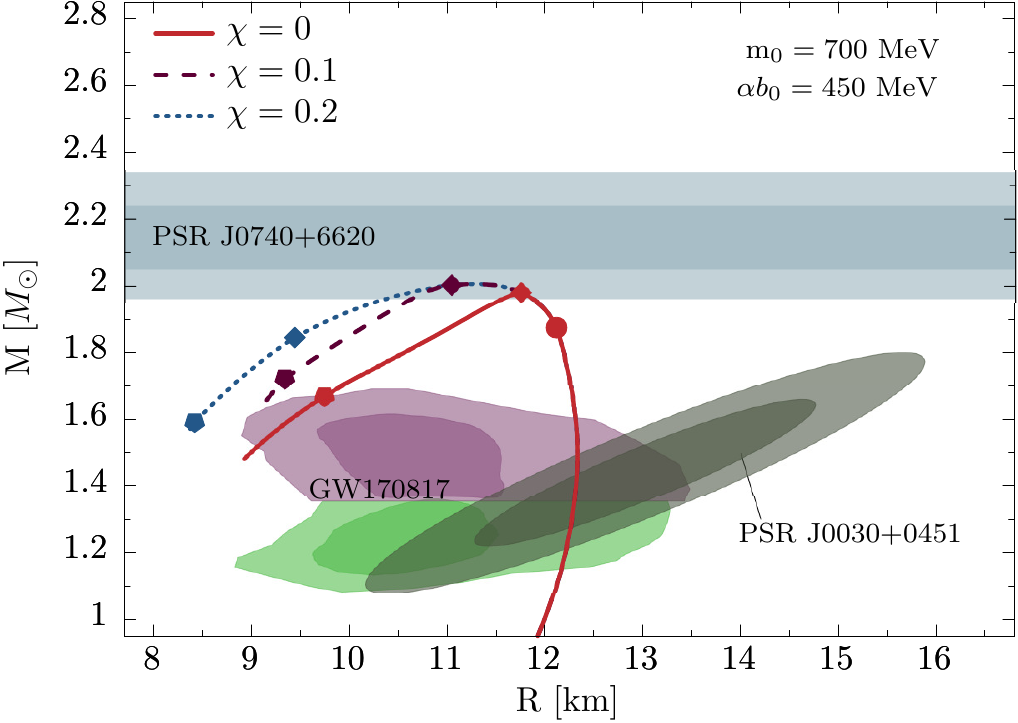}
    \caption{Mass-radius sequences for compact stars as solutions of the TOV equations for $m_0=700$~MeV, for $\alpha b_0=350~$MeV (upper left), $\alpha b_0=350~$MeV (upper right), $\alpha b_0=400~$MeV (lower left), $\alpha b_0=450~$MeV (lower right). The circles show the coexistence of the chirally broken and chirally restored phases. The onsets of up and down quarks are marked by pentagons and diamonds, respectively. We note that, in the upper-left figure, for \mbox{$\chi=0.2$}, the onset of up and down quarks is almost simultaneous. The inner (outer) gray band shows the 68.3\%~(95.4\%) credibility regions for the mass of PSR J0740+6620~\citep{Cromartie:2019kug}. The inner (outer) green and purple bands show 50\%~(90\%) credibility regions obtained from the recent GW170817~\citep{Abbott:2018exr} event for the low- and high-mass posteriors. The inner (outer) black region corresponds to the mass and radius constraint at 68.2\% (95.4\%) obtained for PSR J0030+0451 by the group analyzing NICER X-ray data~\citep{Miller:2019cac}.}
    \label{fig:m_r}
  \end{center}
  \end{figure*}

\section{Properties of compact stars}
\label{sec:mass_radius}

  \begin{table}[t!]\begin{center}\begin{tabular}{|c||c|c|c|c|}
      \hline        & \multicolumn{4}{c|}{$\alpha b_0~$ [MeV]}                              \\ \hline
      $\chi$        & $350$           & $370$           & $400$           & $450$           \\ \hline\hline
      $0$           & $2.11$, $11.89$ & $2.06$, $11.72$ & $2.01$, $11.64$ & $1.98$, $11.74$ \\ \hline
      $0.1$         & $2.13$, $11.43$ & $2.08$, $11.22$ & $2.03$, $11.11$ & $2.01$, $11.27$ \\ \hline
      $0.2$         & $2.13$, $11.43$ & $2.08$, $11.22$ & $2.03$, $11.11$ & $2.01$, $11.27$ \\ \hline
      \end{tabular}\end{center}
      \caption{Maximal neutron-star masses in units of $M_\odot$ and corresponding radius in km (separated by comma) for different values of $\alpha b_0$ and $\chi$ parameters for $m_0=700~$MeV.}
      \label{tab:max_masses}
  \end{table}

  In this section, we explore the physics of dynamical sequential phase transitions at high baryon density on the structure of neutron stars. 
      
  \subsection{TOV solutions}

    We used the EoSs 
    introduced in the previous section to solve the general-relativistic Tolman-Oppenheimer-Volkoff (TOV) equations~\citep{Tolman:1939jz, Oppenheimer:1939ne} for spherically symmetric objects, 
    \begin{subequations}\label{eq:TOV_eqs}
    \begin{align}
       \frac{\dd P(r)}{\dd r} &= -\frac{\left(\epsilon(r) + P(r)\right)\left(M(r) + 4\pi r^3 P(r)\right)}{r \left(r-2M(r)\right)} \textrm,\\
       \frac{\dd M(r)}{\dd r} &= 4\pi r^2 \epsilon(r)\textrm,
    \end{align}
    \end{subequations}
    with the boundary conditions \mbox{$P(r=R) = 0$}, \mbox{$M = M(r=R)$}, where $R$ and $M$ are the radius and the mass of a neutron star, respectively. Once the initial conditions are specified based on a given equation of state, namely the central pressure $P_c$ and the central energy density $\epsilon_c$, the internal profile of a neutron star can be calculated.

    In general, there is one-to-one correspondence between the  EoS and the \mbox{mass-radius} relation calculated  from Eqs. (\ref{eq:TOV_eqs}). In Fig.~\ref{fig:m_r}, we show the relationship of mass versus radius, for the calculated sequences of compact stars for $m_0=700~$MeV, for $\alpha b_0=350~$MeV (upper left), $\alpha b_0=370~$MeV (upper right), $\alpha b_0=400~$MeV (lower left), and $\alpha b_0=450~$MeV (lower left). In each panel, three calculated sequences for \mbox{$\chi=0$} (red: solid line), \mbox{$\chi=0.1$} (purple: dashed line), and \mbox{$\chi=0.2$} (blue: dotted line) are shown. Also shown are the state-of-the-art constraints: the high-precision mass measurement of the high-mass pulsar PSR~J0740+6620~\citep{Cromartie:2019kug}, constraints from two recent GW170817~\citep{Abbott:2018exr} events, and the mass-radius constraint obtained for PSR J0030+0451 by the group analyzing NICER X-ray data~\citep{Miller:2019cac}. We note that the \mbox{mass-radius} relations obtained in the hybrid QMN model remain in good agreement with the aforementioned constraints. In each curve the onsets of down and up quarks are marked with diamonds and pentagons, respectively.

    Because the introduction of the quark-vector repulsive coupling shifts the onset of quarks to higher densities, when compared to the case with vanishing coupling (see Fig.~\ref{fig:p_n}), all three sequences in each of the panels share a chiral phase transition in the high-mass part (red circle), but still below the \mbox{$2~M_\odot$} constraint, at around $1.8~M_\odot$. We note that increasing the value of $\alpha b_0$ softens the chiral transition, which eventually becomes a smooth crossover for $\alpha b_0 = 450~$MeV. We emphasize that an abrupt change in a mass-radius profile in the high-mass part of the sequence is, in general, a result of a phase transition. It leads to a softening of the EoS, meaning that it is accompanied by a rapid flattening of the mass-radius sequence. This is vividly seen in the case of the chiral phase transition on the mass-radius curves in all panels of  Fig.~\ref{fig:m_r}. In a model with sequential transitions, a rapid flattening of the mass-radius curves is not necessarily associated with the deconfinement transition, and hence does not dictate the existence of quark matter in the core of a neutron star, in contrast to previous studies~\citep{Alvarez-Castillo:2016wqj}. Interestingly, the chiral phase transition occurs at masses and radii accessible in the GW190425 merger event~\citep{Abbott:2020uma}. We note that this is similar to the proton flow constraint from HIC in the isospin-symmetric matter and the Hebeler constraint (cf. inset plots in Fig.~\ref{fig:p_n}).

    The extension of the hadronic branch of an EoS due to a finite value of the quark vector-interaction $\chi$ is also reflected in the corresponding mass-radius sequence. Interestingly, the maximal mass is always reached within the hadronic branch of the sequence. For \mbox{$\chi=0$}, this happens just before the density jump associated with the onset of down quark is reached. This is the case, except for $\alpha b_0=370$ and $400~$MeV, where tiny fractions of down quarks appear just before the maximal mass is reached (see the top panel of Fig.~\ref{fig:composition}). The appearance of the down quark makes the matter too soft to sustain from the gravitational collapse. For \mbox{$\chi=0.1$} and $0.2$, the hadronic branch extends beyond the density at which the maximal mass is reached and becomes gravitationally unstable. This is the case for all values of $\alpha b_0$. Eventually, when down and up quark are sequentially populated, the matter is still not stiff enough to sustain from the collapse and turn into an additional family of stable hybrid compact stars. This is shown in Fig.~\ref{fig:composition}, for $m_0=700~$MeV and $\alpha b_0 = 370~$MeV, where for different values of $\chi$, the densities at which the maximal mass is reached are marked by gray solid vertical lines. Clearly, the density for \mbox{$\chi=0.1$} (middle panel) is greater than for $\chi=0$ (top panel), and the hadronic branch extends beyond it. For $\chi=0.2$, the hadronic branch extends even further, however the maximal mass stays the same. Thus, we conclude that a further increase of the quark-vector coupling does not support the maximal-mass constraint. In Table~\ref{tab:max_masses}, we show the values of the maximal masses of a neutron star and corresponding radii obtained in each parametrization.

    We stress that the above structure persists to large extent when the chiral invariant mass $m_0$ changes. In order for the pure quark matter to be present in the cores of neutron stars, the repulsive interactions among them should be strong. On the other hand, the self-consistent treatment of the interactions in the hybrid QMN model also shifts the onset of the quark matter in the stellar sequence to higher densities. This eventually prevents the neutron stars with quark matter cores from existence in the gravitationally stable branch of the sequence. This leads to a general conclusion that the existence of hybrid stars is rather excluded in the current hybrid QMN model.

  \subsection{Tidal deformability}

    The dimensionless tidal deformability parameter $\Lambda$ can be computed through its relation to the Love number $k_2$~\citep{Hinderer:2007mb,Damour:2009vw,Binnington:2009bb,Yagi:2013awa,Hinderer:2009ca},
    \begin{equation}
       \Lambda = \frac{2}{3} k_2 C^{-5} \textrm,
    \end{equation}
    where $C = M/R$ is the star compactness parameter, with $M$ and $R$ being the total mass and radius of a star. The Love number $k_2$ reads
    \begin{equation}
    \begin{split}
       k_2 &= \frac{8C^5}{5} \left(1-2C\right)^2 \left[ 2+2C(y-1)-y \right] \times \\
          \times & \Big(2C\left[6- 3y + 3C(5y-8) \right]\\
          +      & 4C^3[13-11y+C(3y-2)+2C^2(1+y)] \\
          +      & 3(1-2C)^2[2-y+2C(y-1)\ln{(1-2C)}]\Big)^{-1} \textrm,
    \end{split}
    \end{equation}
    where $y = R\beta(R)/H(R)$. The functions $H(r)$ and $\beta(r)$ are given by the following set of differential equations:
    \begin{eqnarray}
    \frac{\dd \beta}{\dd r}&=&2 \left(1 - 2\frac{M(r)}{r}\right)^{-1} \nonumber\\
    && H\left\{-2\pi \left[5\epsilon(r)+9 P(r)+\frac{\dd \epsilon}{\dd P}(\epsilon(r)+P(r))\right]\phantom{\frac{3}{r^2}} \right. \nonumber\\
    && \left.+\frac{3}{r^2}+2\left(1 - 2\frac{M(r)}{r}\right)^{-1} \left(\frac{M(r)}{r^2}+4\pi r P(r)\right)^2\right\}\nonumber\\
    &&+\frac{2\beta}{r}\left(1 - 2\frac{M(r)}{r}\right)^{-1}\nonumber\\
    &&  \left\{\frac{M(r)}{r}+2\pi r^2 (\epsilon(r)-P(r)) - 1\right\}~\textrm,\\
    \frac{\dd H}{\dd r}&=& \beta \textrm.
    \end{eqnarray}
    The above equations have to be solved along with the TOV equations~(\ref{eq:TOV_eqs}). The initial conditions are \mbox{$H(r\rightarrow 0) = c_0 r^2$} and \mbox{$\beta(r\rightarrow 0) =2c_0 r$}, where $c_0$ is a constant, which is irrelevant in the expression for the Love number $k_2$.

    \begin{figure}[t]
    \begin{center}
      \includegraphics[width=0.95\linewidth]{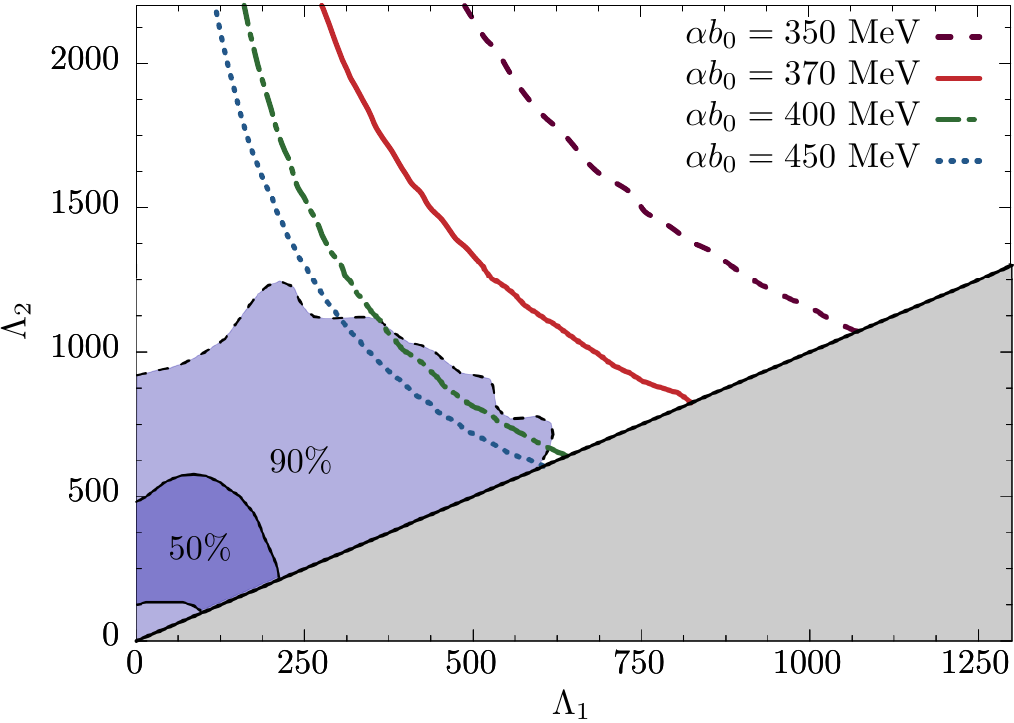}
      \caption{Relation between tidal deformabilities $\Lambda_1$ and $\Lambda_2$ of two compact stars that merged in the GW170817 event~\citep{Abbott:2018exr}. For comparison, also shown are the 50\% and 90\% fidelity regions from the analysis of the GW signal by the LIGO-VIRGO collaboration~\citep{Abbott:2018exr}. The gray shading corresponds to the unphysical region $\Lambda_2 < \Lambda_1$.}
      \label{fig:L1_L2}
    \end{center}
    \end{figure}

    In Fig.~\ref{fig:L1_L2}, we plot the tidal deformability parameters $\Lambda_1$ vs $\lambda_2$ of the high- and low-mass members of the binary merger together with the 50\% and 90\% fidelity regions obtained by the LVC analysis of the GW170817 event~\citep{Abbott:2018exr}. Because the extracted most probable masses of the members are below the mass where the chiral phase transition occurs (cf. Fig.~\ref{fig:m_r}), the $\Lambda_1$ vs. $\Lambda_2$ diagrams are the same for given $m_0$ and $\alpha b_0$, irrespective of the value of the parameter $\chi$. We note that the tidal deformability parameter requires sufficiently soft EoS around the saturation density. This is seen in the figure, where the smallest tidal deformability and thus the best agreement with the constraint is obtained for the highest value of $\alpha b_0=450~$MeV, which corresponds to the softest EoS at low density. We remark that this is in similar fashion to the proton flow constraint. On the other hand, the $2~M_\odot$ requires a sufficiently stiff EoS at higher densities. In Table~\ref{tab:max_masses}, we show the maximal masses obtained for each set of parameters. Inversely to the tidal deformability, the most massive stars are obtained for the stiffest EoS, namely for $\alpha b_0=350~$MeV. In general, the two constraints are of exclusive character. The interplay between them could be further used to fix the allowed range of external model parameters. This would be of particular use in a class of effective models in which the low- and high-density regimes are not treated independently, but rather combined in a consistent unified framework.
  
\section{Conclusions}
\label{sec:conclusions}

  Despite the success of LQCD at a finite temperature and vanishing density, the nature of the chiral and deconfinement phase transitions at a low temperature and high density remains unknown. At present, the investigation of matter under such extreme conditions is almost exclusively considered within either purely hadronic or purely quark effective models, with only few approaches to bridge the gap between them. They are typically based on excluded volume approach or a Polyakov loop representing a phenomenological description of the transition from the confined to the deconfined phase~\citep{Steinheimer:2011ea, Berges:1998ha, Meyer:1999bx, Lawley:2006ps, Dexheimer:2009hi}.

  In this work, we utilized a superior approach to any previous modeling, the hybrid QMN model~\citep{Benic:2015pia}, to quantify  the EoS of cold and dense nuclear matter. This model goes beyond the common two-phase approach to the EoS and embeds the interplay between the quark confinement and the chiral symmetry breaking in a dynamical way into a single unified framework. We have extended the previous version of the model by including repulsive interactions among quarks mediated by the exchange of vector mesons. Within this approach, we systematically investigated the EoS of cold and dense isospin-symmetric matter, as well as asymmetric  matter under NS conditions. We constructed the mass-radius relations based on solutions of the TOV equations.

  We find that the class of EoSs obtained in the considered model  feature sequential chiral and deconfinement phase transitions. We demonstrate that, as a consequence of the realization of the chiral symmetry restoration by parity doubling within the hadronic phase, a strong first-order phase transition invalidates the implication that a flattening, eventually even occurrence of a mass-twin phenomenon, of the mass-radius relation for compact stars at $2~M_\odot$, could inevitably signal the  deconfinement phase transition in compact stars~\citep{Alvarez-Castillo:2016wqj}. We verified that such  flattening  appears due to  the chiral phase transition which also lies  within the region of the proton flow constraint in isospin-symmetric matter~\citep{Danielewicz:2002pu}, as well as within the constraint derived in Ref.~\cite{Hebeler:2013nza} for the matter under NS conditions. Moreover, the chiral transition is featured in the mass region accessible by the recent GW190425 merger event~\citep{Abbott:2020uma}, that is, $1.46-1.87~M_\odot$, which is of particular interest from the observational perspectives.

  The EoS obtained in the hybrid QMN model, under the NS conditions, is comprised of four consecutive density regions, from low to high density: confined and chirally broken phase, confined and chirally restored phase, partially deconfined, and deconfined phase. In the intermediate phases, the confined and deconfined matter may coexist as what  could be a possible manifestation  of the conjectured quarkyonic phase~\citep{Hidaka:2008yy, McLerran:2008ua, Andronic:2009gj, McLerran:2008ua, McLerran:2018hbz, Jeong:2019lhv}. We stress that such a sequential structure of an EoS cannot be obtained in a class of pure quark models or even in a class of hadron-quark models, where the confined and deconfined phases are treated independently. Recent studies in this direction provide an interesting scenario for a rapid increase in the pressure within densities achieved in NSs~\citep{McLerran:2018hbz, Jeong:2019lhv}. It would also  be interesting to establish the cooling and transport properties of the hybrid QMN model EoS, especially the mixed phase comprised of nucleons and down quarks, as well as to establish their influence on the thermal and rotational evolution of compact objects.

  We embedded the repulsive quark-vector interactions in the QMN model and studied their  consequences for the phenomenological description of compact stellar objects. In particular, we analyzed a possible occurrence of a quark matter in the NS core. The inclusion of the quark-vector repulsion implies  a shift of the onset of quarks to higher densities. This results  in the mass-radius diagram with the maximal mass around $2~M_\odot,$ which is within the chirally restored hadronic configuration. 
  
  In the present studies, the transition to pure quark matter is likely to appear in the part of the stellar sequence that is already gravitationally unstable. We note, however, that the presence of the attractive diquark interaction can induce scalar diquark condensation in the quark phase and reduce the density of the onset of the quark matter~\citep{Klahn:2006iw}. This comes with a softening of the EoS which, however, can be compensated by a stronger vector meson coupling~\citep{Klahn:2013kga,Alvarez-Castillo:2018pve}. Moreover, in  recent LQCD studies, it was  found that parity doubling also occurs in the hyperon channels~\citep{Aarts:2018glk}. The inclusion of heavier flavors is  known to soften  the equation of state. Therefore, it would be of further  interest to  consider  the  QMN  equations of state that include the attractive diquark interaction and the hyperon degrees of freedom in order to  test a model's compliance with the $2~M_\odot$ constraint. Work in this direction  is already  in progress. 

  We show how modern astrophysical constraints on the maximum-mass \citep{Cromartie:2019kug}, the tidal deformability from the binary merger GW170817~\citep{Abbott:2018exr}, and recent simultaneous mass-radius constraints from the NICER experiment~\citep{Riley:2019yda, Miller:2019cac}, compiled together, allow for a consistent determination of parameters characterizing the EoS of dense nuclear matter. We demonstrate that not only too soft (excluded by the maximum mass constraint) but also too stiff (the tidal deformability constraint) equations of state may be ruled out in the current approach. With the optimal choice of parameters in the QMN model, we introduced a class of EoSs that fulfill the above astrophysical constraints, as well as phenomenological constraints imposed by flow data  from heavy ion collisions. This makes the present hybrid QMN model EoSs superior to previously proposed class of EoSs obtained in pure-hadronic or pure-quark models, as well as in  hybrid models based on a two-phase approach. It would be of interest to apply a Bayesian analysis method for selecting the most probable equation of state under a set of modern constraints from compact star physics, and from data obtained in heavy ion collisions. For applications of the QMN model EoS to multi-messenger astronomy in simulations of merger or supernova events as well as heavy ion collision, it is necessary to perform a finite-temperature extension. This is a straightforward task in the presented field-theoretical model as first results for corresponding phase diagrams  of the unified quark-hadron matter description convincingly demonstrate~\citep{Marczenko:2018jui,Marczenko:2019trv}.

  We provided a first step toward a unified quark-hadron matter EoS for multi-messenger astronomy that allows conclusions for the composition of neutron star interiors and the phase structure of matter under extreme conditions, in accordance with mass-radius measurements of compact stars and heavy ion collision phenomenology. The next steps include be the generalization of the QMN model to include color superconducting quark matter phases that may allow the existence of hybrid quark-hadron stars due to the lowering of the onset mass for deconfinement while fulfilling the maximum mass constraint for a sufficiently large coupling to vector meson fields.

\begin{acknowledgements}
       This work was partly supported by the Polish National Science Center (NCN), under OPUS Grant No. 2018/31/B/ST2/01663 (K.R. and C.S.), Opus Grant~No. UMO-2019/33/B/ST9/03059 (D.B.), and Preludium Grant No. UMO-2017/27/N/ST2/01973 (M.M.). D.B. received support from the National Research Nuclear University (MEPhI) within the Russian Academic Excellence Project under contract No. 02.a03.21.0005. K.R. also acknowledges stimulating discussions with Larry McLerran and the support of  the  Polish  Ministry  of  Science and Higher Education.

\end{acknowledgements}

\bibliography{paper} 

\begin{appendix} 
  \section{Impact of statistical confinement on equation of state}
  \label{sec:appendixA}
    
    In this appendix, we provide an example of the stiffening of the EoS due to the statistical confinement in the hybrid QMN model. For illustration, let us consider two EoSs. For the first EoS, we take a non-interacting gas of massless fermions at zero temperature. In this case, the pressure, $P_1(\mu)$, can be evaluated analytically as a function of chemical potential $\mu$, as
    \begin{equation}\label{eq:a1}
      P_1(\mu) = - \int\limits_0^\mu \frac{\dd p}{2\pi^2}\;p^2 \left(p - \mu\right)= \frac{\mu^4}{24\pi^2} \textrm,
    \end{equation}
    where the upper integration limit is due to the Fermi-Dirac distribution function for a massless particle in the zero-temperature limit, $\theta\left(\mu - p\right)$. For the second EoS, let us consider the modification of the distribution function as in Eq.~\eqref{eq:cutoff_nuc}. For simplicity, we also set $\alpha=1$. The pressure, $P_2(\mu)$, is then obtained as
   \begin{equation}\label{eq:a2}
      P_2(\mu) = - \int\limits_0^\mu \frac{\dd p}{2\pi^2}\;p^2 \left(p - \mu\right)\theta\left(b - p\right) = 
      \begin{cases} 
      \frac{\mu^4}{24\pi^2} \textrm{, if } b \geq \mu\\
      \frac{b^4}{24\pi^2} \textrm{, if } b < \mu
      \end{cases}\textrm.
    \end{equation}
    Thus, if $b \geq \mu$, the pressure $P_2(\mu)$ evaluates to $P_1(\mu)$. However, if $b < \mu$, the mechanism of statistical confinement sets in and the integration region is limited up to $b$, and pressure $P_2(\mu)  < P_1(\mu)$. Let us assume that in the region where $b < \mu$, the functional form of the $b$ field is $b=\gamma \mu$, where $\gamma$ is a positive constant, such that $\gamma \in (0, 1)$. Then we get the following:
    \begin{equation}
      P_2(\mu) = \frac{b^4}{24\pi^2} = \gamma^4 \frac{\mu^4}{24\pi^2} = \gamma^4 P_1(\mu)  < P_1(\mu)\textrm.
    \end{equation}
    Let us consider two values of chemical potential, $\mu_1$ < $\mu_2$, such that $P_1(\mu_1) = P_2(\mu_2)$. From this, it follows that $\mu_1^4 = \gamma^4\mu_2^4$. Because the density is defined as the derivative of pressure w.r.t chemical potential, $\rho = \partial P / \partial \mu$, we get
    \begin{equation}
     \rho_2(\mu_2) = \gamma^4\frac{\mu_2^3}{6\pi^2} = \frac{\mu_1}{\mu_2}\frac{\mu_1^3}{6\pi^2} = \frac{\mu_1}{\mu_2}\rho_1(\mu_1) < \rho_1(\mu_1) \textrm.
    \end{equation}
   The last inequality holds by the assumption that $\mu_1 < \mu_2$. Thus, the pressure $P_2$ increases faster than $P_1$ as a function of density $\rho$. This is the stiffening effect observed in the class of EoSs obtained in the hybrid QMN model. We note that, conversely to the case discussed in this appendix, the appearance of quarks always provides an extra contribution to the pressure, thus, softening the corresponding EoS.

\end{appendix}

\end{document}